%% file: main.tex
\documentclass[sigconf]{acmart}
\usepackage{multirow}
\usepackage{textcomp}
\usepackage{balance}
\usepackage{amsmath}
\usepackage{amssymb}
\usepackage{amsfonts}
\usepackage{caption}
\usepackage[labelformat=simple]{subcaption}
\usepackage{graphicx}
\usepackage{bbm}
\usepackage{gensymb}
\usepackage{enumitem}
\setlist{nolistsep}
\usepackage{subfiles}
\usepackage{url}
\usepackage[export]{adjustbox}
\usepackage[]{natbib}
\usepackage{epstopdf}
\usepackage{algorithmic}
\usepackage{hyphenat}
\usepackage[ruled,vlined]{algorithm2e} 
\usepackage{booktabs,subcaption,amsfonts,dcolumn}
\usepackage{fancyhdr,amsmath,amssymb}

\renewcommand{\algorithmicensure}{\textbf{Output:}}

\newcolumntype{d}[1]{D..{#1}}

\copyrightyear{2019}
\acmYear{2019}
\setcopyright{acmlicensed}
\acmConference[BuildSys '19]{BuildSys '19: ACM Conference on Systems for Energy-Efficient Buildings, Cities, and Transportation}{November 13--14, 2019}{New York, NY}
\acmBooktitle{BuildSys '19: ACM Conference on Systems for Energy-Efficient Buildings, Cities, and Transportation, November 13--14, 2019, New York, NY}
\acmPrice{15.00}
\acmDOI{xx.xxxx/xxxxxxx.xxxxxxx}
\acmISBN{978-1-xxxx-xxxx-x/xx/xx}

\def\BibTeX{{\rm B\kern-.05em{\sc i\kern-.025em b}\kern-.08emT\kern-.1667em\lower.7ex\hbox{E}\kern-.125emX}}

\title[Multi-Scale, Cascaded RNNs for Radar Classification]{One Size Does Not Fit All: \\ Multi-Scale, Cascaded RNNs for Radar Classification}

\author[D. Roy]{Dhrubojyoti Roy}
\authornote{Both authors contributed equally to this research.}
\email{roy.174@osu.edu}
\author[S. Srivastava]{Sangeeta Srivastava}
\authornotemark[1]
\email{srivastava.206@osu.edu}
\affiliation{%
  \institution{The Ohio State University}
}

\author[A. Kusupati]{Aditya Kusupati}
\email{kusupati@cs.washington.edu}
\affiliation{%
  \institution{University of Washington}
  \institution{Microsoft Research India}
}

\author[P. Jain]{Pranshu Jain}
\email{anz178419@cse.iitd.ac.in}
\affiliation{%
  \institution{Indian Institute of Technology Delhi}
}

\author[M. Varma]{Manik Varma}
\email{manik@microsoft.com}
\affiliation{%
  \institution{Microsoft Research India}
  \institution{Indian Institute of Technology Delhi}
}

\author[A. Arora]{Anish Arora}
\email{arora.9@osu.edu}
\affiliation{%
 \institution{The Ohio State University}
  \institution{The Samraksh Company}
}



\def\vec#1{\mathchoice%
	{\mbox{\boldmath $\displaystyle\bf#1$}}
	{\mbox{\boldmath $\textstyle\bf#1$}}
	{\mbox{\boldmath $\scriptstyle\bf#1$}}
	{\mbox{\boldmath $\scriptscriptstyle\bf#1$}}}

\begin{document}
\subfile{abstract}
\maketitle

\subfile{introduction}

\subfile{relatedwork}
\subfile{sysmodel}

\subfile{solarch}

\subfile{evaluation}
\subfile{discussion}
\subfile{conclusion}

\subfile{acknowledgement}
\bibliographystyle{ACM-Reference-Format}
\bibliography{main}
\subfile{appendix}
\end{document}

%% file: abstract.tex
\begin{abstract}
Edge sensing with micro-power pulse-Doppler radars is an emergent domain in monitoring and surveillance with several smart city applications. Existing solutions for the clutter versus multi-source radar classification task are limited in terms of either accuracy or efficiency, and in some cases, struggle with a trade-off between false alarms and recall of sources. We find that this problem can be resolved by learning the classifier across multiple time-scales. We propose a multi-scale, cascaded recurrent neural network architecture, MSC-RNN, comprised of an efficient multi-instance learning (MIL) Recurrent Neural Network (RNN) for clutter discrimination at a lower tier, and a more complex RNN classifier for source classification at the upper tier. By controlling the invocation of the upper RNN with the help of the lower tier conditionally, MSC-RNN achieves an overall accuracy of 0.972. Our approach holistically improves the accuracy and per-class recalls over ML models suitable for radar inferencing. Notably, we outperform cross-domain handcrafted feature engineering with time-domain deep feature learning, while also being up to $\sim$3$\times$ more efficient than a competitive solution.
\end{abstract}

%% file: introduction.tex
\section{Introduction}
\label{sec_intro}

With the rapid growth in deployment of Internet of Things (IoT) sensors in smart cities, the need and opportunity for computing increasingly sophisticated sensing inferences on the edge has also grown. This has motivated several advances in designing resource efficient sensor inferences, particularly those based on machine learning and especially deep learning. The designs, however, encounter a basic tension between achieving efficiency while preserving predictive performance that motivates a reconsideration of state-of-the-art techniques.  

In this paper, we consider a canonical inference pattern, namely discriminating clutter from several types of sources, in the context of a radar sensor. This sort of $N\!+\!1$--class classification problem, where $N$ is the number of source types, has a variety of smart city applications, where diverse clutter is the norm. These include triggering streetlights smartly, monitoring active transportation users (pedestrians, cyclists, and scooters), crowd counting, assistive technology for safety, and property surveillance. As an example, streetlights should be smartly triggered on for pedestrians but not for environmental clutter such as trees moving in the wind. Similarly, property owners should be notified only upon a legitimate intrusion but not for passing animals.

\begin{figure}[t]
	\centering
	\begin{subfigure}[t]{0.20\textwidth}
		\includegraphics[width=\textwidth]{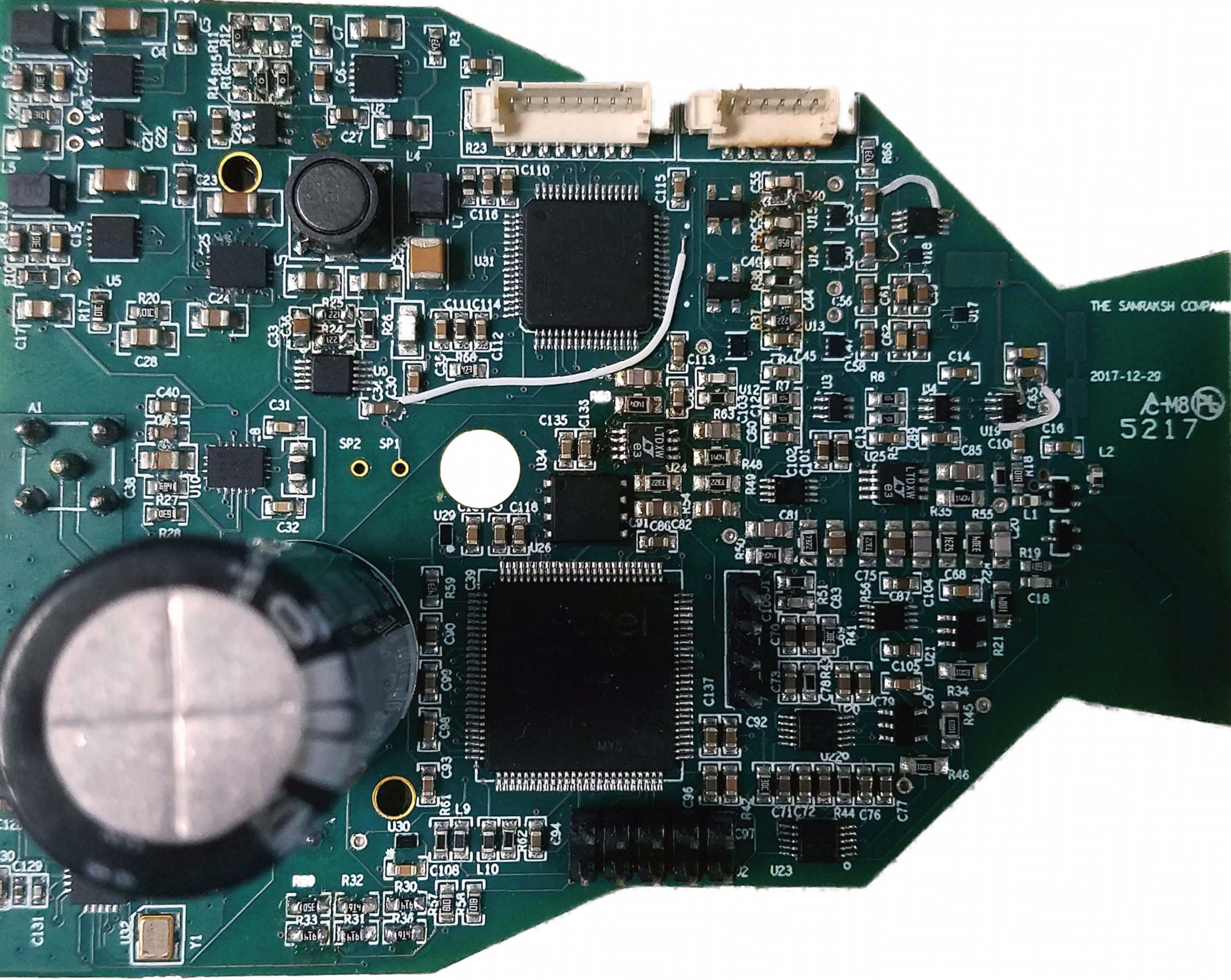}
		\caption{Micro-power PDR system}
	\end{subfigure}%
	~
	%
	\begin{subfigure}[t]{0.23\textwidth}
		\includegraphics[width=\textwidth]{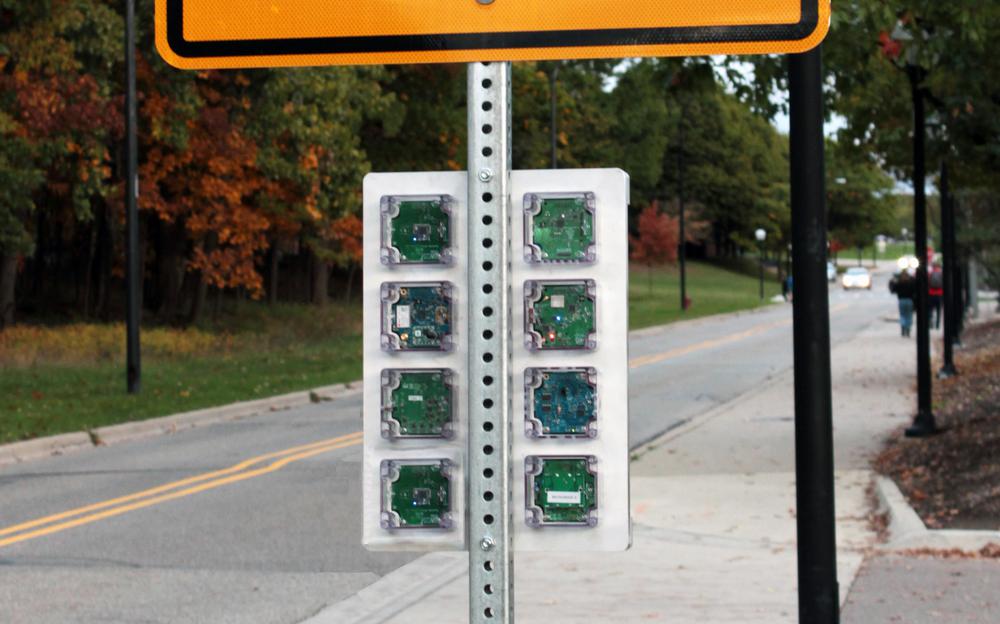}
		\caption{Solar harvested Signpost platform supporting low power sensors \cite{adkins2018signpost}}
	\end{subfigure}
	\caption{The micro-power pulse-Doppler radar (PDR) device can be independently deployed or interfaced with existing multi-sensor smart city platforms such as Signpost (figure adapted from \cite{adkins2018signpost}, Copyright \textcopyright 2019 ACM, Inc.)}
	\label{fig_radar}
\end{figure}

The radar is well suited in the smart city context as it is privacy preserving in contrast to cameras. Moreover, it consumes low power ({\small{$\sim$}}15mW), because of which it can be deployed at operationally relevant sites with little dependence on infrastructure, using, for instance, a small panel solar harvester or even a modest sized battery, as shown in Figure \ref{fig_radar}. Experiences with deploying sensors in visionary smart city projects such as Chicago's Array of Things \cite{catlett2017array, AoT} and Sounds of New York City \cite{Bello2019} have shown that wired deployments on poles tend to be slow and costly, given constraints of pole access rights, agency coordination, and labor unions, and can sometimes be in suboptimal locations. Using a low-power sensor that is embedded wirelessly or simply plugged in to existing platforms while imposing only a nominal power cost simplifies smart city deployment.

Table \ref{tab_tradeoff} illustrates an efficiency-accuracy trade-off for the canonical inference pattern with $N=2$, wherein clutter is distinguished from human and other (i.e., non-human) sources. The more accurate deep models, the Convolutional Neural Network (CNN) \cite{krizhevsky2012imagenet} and the Long Short-Term Memory (LSTM) \cite{hochreiter1997long}, that we machine-learned for this 3-class classifier from a reference dataset are significantly less efficient, in terms of speed and therefore power consumption.  In contrast, the more efficient shallow solution, Support Vector Machine (SVM), is significantly less accurate. While the SVM classifier has been implemented to operate in near real-time on the Cortex-M3 single-microcontroller processor in the device depicted in Fig.~\ref{fig_radar}(a), neither the CNN nor the LSTM per se yield a near real-time implementation. To implement deep models in near real-time on the M3, we therefore consider model optimization with recent state-of-art-techniques such as fast gated RNNs (FastGRNN) \cite{kusupati2018fastgrnn} and Early-exit Multi-Instance RNNs (EMI-LSTM and EMI-FastGRNN) \cite{dennis2018multiple}.  However, Table \ref{tab_tradeoff} illustrates that the trade-off remains: the best accuracy we achieve, namely with the FastGRNN, has significantly lower efficiency than the best efficiency achieved, namely with EMI-FastGRNN, but that has an accuracy that is comparatively significantly worse.

\begin{table}[t]
	\vspace*{4.5mm}
	\caption{Trade-offs in accuracy and runtime efficiency for the 3-class radar problem (window length 1s, feature computation overhead ignored for SVM, dataset and machine architecture details are in Section 5)}
	\begin{tabular}{l|crc}
		\hline
		\multicolumn{1}{c|}{\textbf{ML Model}} & \multicolumn{1}{c}{\textbf{Accuracy}} & \multicolumn{1}{c}{\textbf{FLOPS}} & \multicolumn{1}{c}{\textbf{Real-time?}} \\ \hline\hline
		SVM (15 features)   & 0.85  &    37K 	& Yes     \\
		LSTM                & 0.89  &    100K   & No      \\ 
		CNN (1s FFT)        & 0.91  &    1.3M   & No      \\ 	
		\hline
		EMI-LSTM            & 0.90  &    20K   & Yes     \\
		FastGRNN            & 0.96  &    35K   & Yes     \\	 
		EMI-FastGRNN        & 0.88  &     8K   & Yes     \\
		\hline
	\end{tabular}
	\label{tab_tradeoff}
\vspace*{-3mm}
\end{table}

\vspace{1mm}
\noindent
\textbf{Problem Statement.}\enspace In this work, we investigate alternative optimizations of deep models for the above classification task that achieve both high accuracy and speed.  In doing so, we do not wish to sacrifice the recall performance for achieving high precision.  For instance, radar sensing applications require that the clutter recall be very high so that there are minimal false alarms. However, a solution that restricts false alarms at the cost of detectability (i.e., low source recall, where a source could be either human or non-human) would be undesirable as it would have limited applicability in the smart city contexts discussed above. 

\vspace{1mm}
\noindent
\textbf{Solution Overview.}\enspace The $N\!+\!1$-class radar problem, where the $+1$-class is clutter, conflates discrimination between classes that are conceptually different. In other words, discriminating clutter from sources has a different complexity from that of disambiguating source types. This insight generalizes when the sources themselves are related by a hierarchical ontology, wherein different levels of source types involve concepts of correspondingly different complexity of discrimination.  

By way of example, in the 3-class clutter vs.~human vs.~non-human classification problem, discriminating clutter from sources turns out to be simpler than discriminating the more subtle differences between the source types. 
Using the same machine architecture for 3 classes of discrimination leads to the accuracy-efficiency trade-off, as the last two rows of Table \ref{tab_tradeoff} indicate. A more complex architecture suffices for discriminating among source types accurately, whereas a simpler architecture more efficiently suffices for discriminating clutter from sources, but hurts the accuracy of discriminating between source types.  

We, therefore, address the problem at hand with an architecture that decomposes the classification inference into different hierarchical sub-problems.  For the 3-class problem, these are:  (a) Clutter vs Sources, and (b) Humans vs.~Non-humans \textit{given} Sources. For each sub-problems we choose an appropriate learning architecture; given the results of Table \ref{tab_tradeoff}, both architectures are forms of RNN albeit {\em with learning at different time-scales}.  The lower tier RNN for (a) uses a short time-scale RNN, the Early-exit Multi-Instance RNN (EMI-FastGRNN) \cite{dennis2018multiple, kusupati2018fastgrnn}, whereas the higher tier for (b) uses a longer time-scale RNN, a FastGRNN \cite{kusupati2018fastgrnn}, which operates at the level of windows (contiguous, fixed-length snippets extracted from the time-series) as opposed to short instances within the window. The upper tier uses the features created by the lower tier as its input; for loss minimization, both tiers are jointly trained. To further improve the efficiency, we observe that source type discrimination needs to occur only when a source is detected and clutter may be the norm in several application contexts. Hence, the less efficient classifier for (b) is invoked only when (a) discriminates a source: we refer to this as cascading between tiers. The joint training loss function is refined to emulate this cascading.
We call this architecture {\em Multi-Scale, Cascaded RNNs (MSC-RNN)}.

\vspace{1mm}
\noindent
\textbf{Contributions.}\enspace 
Our proposed architecture exploits conditional inferencing at multiple time-scales to jointly achieve superior sensing and runtime efficiency over state-of-the-art alternatives. To the best of our knowledge, this approach is novel to deep radar systems. For the particular case of the 3-class problem, MSC-RNN performs as follows on the Cortex-M3: \\

\begin{tabular}{ccccc}
\hline
\multicolumn{1}{c}{\textbf{Accuracy}} & 
\multicolumn{1}{c}{\textbf{\begin{tabular}[c]{@{}c@{}}Clutter\\ Recall\end{tabular}}} &
\multicolumn{1}{c}{\textbf{\begin{tabular}[c]{@{}c@{}}Human\\ Recall\end{tabular}}} &
\multicolumn{1}{c}{\textbf{\begin{tabular}[c]{@{}c@{}}Non-human\\ Recall\end{tabular}}} &
\multicolumn{1}{c}{\textbf{FLOPS}}
\\ \hline\hline
0.972  &  1  & 0.92 & 0.967 &  9K
\\ \hline
\end{tabular}
\label{tab_overview}
\vspace{1.5mm}

\ \\
\noindent
Its accuracy and per-class recalls are mostly better than, and in remaining cases competitive with, the models in Table \ref{tab_tradeoff}. Likewise, its efficiency is competitive with that of EMI-FastGRNN, the most efficient of all models, while substantially outperforming it in terms of sensing quality. We also validate that this MSC-RNN solution is superior to its shallow counterparts not only comprehensively, but at each individual tier as well. The data and training code for this project are open-sourced at \cite{mscrnncode}.

Other salient findings from our work are summarized as follows:

\begin{enumerate}[topsep=0cm,after={\advance\csname @topsepadd\endcsname by 0cm}]
    \item Even with deep feature learning purely in the time-domain, MSC-RNN  surprisingly outperforms handcrafted feature engineering in the amplitude, time, and spectral domains for the source separation sub-problem. Further, this is achieved with 1.75-3$\times$ improvement in the featurization overhead.
    
    \item The Tier 1 component of MSC-RNN, which classifies legitimate sources from clutter, improves detectability by up to $2.6\times$ compared to popular background rejection mechanisms in radar literature, even when the false alarm rate is controlled to be ultra-low.
    
	\item MSC-RNN seems to tolerate the data imbalance among its source types better than other compared RNN models. In particular, it enhances the non-dominant human recall by up to 20\%, while simultaneously maintaining or improving the dominant non-human recall and overall accuracy.
\end{enumerate}

\vspace{1mm}
\noindent
\textbf{Organization.}\enspace In Section \ref{sec_relwork}, we present related research and outline the basics of micro-power radar sensing in Section \ref{sec_sysmodel}. In Section \ref{sec_solarch}, we detail the various components in our solution and discuss the training and inference pipelines. We provide evaluation and prototype implementation details in Sections \ref{sec_eval} and \ref{sec_disc} respectively. We conclude and motivate future research in Section \ref{sec_conc}.

%% file: relatedwork.tex
\section{Related Work}
\label{sec_relwork}

\noindent
\textbf{Shallow Radar Sensing. }\enspace Micro-Doppler features have been used in myriad applications ranging from classification \cite{he2014mote, fallDetection, kim2008human} to regression \cite{he2014regression}. Most of these applications employ the short-time Fourier transform (STFT) representation for analyzing micro-Doppler signatures. Although shallow classifiers can be computationally cheaper than deep solutions, the spectrogram generation over a sliding window for the STFT incurs significant computational overhead for real-time applications on single microcontroller devices. In order to decrease this overhead for feature extraction, different feature extraction methods like linear predictive coding \cite{lpc}, discrete-cosine coefficients \cite{molchanov2011ground}, log-Gabor filters with principal component analysis \cite{logGabor}, empirical mode decomposition \cite{EMDecomp} have been investigated in the past. We, on the other hand, use a deep learning approach that learns relevant features from raw time-series data, and avoid spectrogram computation altogether. Feature engineering requires sophisticated domain knowledge, is not assured to be efficient per se, and may not transfer well to solutions for other research problems. Moreover, selection of relevant and non-redundant features requires care for sensing to be robust \cite{roy2017cross}. 

\vspace{1mm}
\noindent
\textbf{Deep Radar Sensing. }\enspace In recent years, there has been significant use of deep learning for radar applications. Most works use spectrogram-based input \cite{kim2016gestureCNN, kim2015humandetCNN, fallDetectionDNN} with deep architectures like CNNs/autoencoders. The authors of \cite{uasDetection} digitize the radio receiver's signal and generate a unique spectral correlation function for the Deep Belief Network to learn signatures from. The pre-processing needed in these applications and the resulting model sizes make them unsuitable for single microcontroller devices. We use raw time-series data in conjunction with variants of RNNs to achieve a faster and efficient solution.

\vspace{1mm}
\noindent
\textbf{Efficient RNN. }\enspace The ability of RNNs in learning temporal features has made it ubiquitous in various sequence modeling tasks. RNNs, albeit theoretically powerful, often fail to reach the best performance due to instability in their training resulting from the exploding and vanishing gradient problem (EVGP) \cite{pascanu2013difficulty}. Gated RNNs like LSTM \cite{hochreiter1997long} and GRU \cite{cho2014properties} have been proposed to circumvent EVGP and achieve the desired accuracy for the given task. A drawback of LSTM and GRU is their model size and compute overhead which makes them unattractive for the near real-time single microcontroller implementations. Recently, FastGRNN \cite{kusupati2018fastgrnn} has been proposed to achieve prediction accuracies comparable to LSTM and GRU while ensuring that the learned models are smaller than 10 KB for diverse tasks. Our proposed hierarchical classifier solution is based on this architecture.

\vspace{1mm}
\noindent
\textbf{Multi-Instance Learning and Early Classification. }\enspace MIL is a weakly supervised learning technique that is used to label sub-instances of a window. MIL has found use in applications from vision \cite{wu2015deep} to natural language processing (NLP) \cite{kotzias2014deep}. It enables a reduction in the computational overhead of sequential models like RNNs by localizing the appropriate activity signature in a given noisy and coarsely-labeled time-series data along with early detection or rejection of a signal \cite{dennis2018multiple}. We use it as our lower tier classifier for clutter versus source discrimination.

\vspace{1mm}
\noindent
\textbf{Multi-Scale RNN. }\enspace One of the early attempts to learn structure in temporally-extended sequences involved using reduced temporal sequences \cite{hinton1990mapping} to make detectability over long temporal intervals feasible in recurrent networks \cite{mozer1992, schmidhuber1992learning}. With the resurgence of RNNs, multi-scale RNNs can discover the latent hierarchical multi-scale structure of sequences \cite{chung2016hierarchical}. While they have been traditionally used to capture long-term dependencies, we use it to design a computationally efficient system. We use different scales of temporal windows for the lower and upper tier RNN. By conditioning the upper tier classifier, which works on longer windows and is hence bulkier we make sure that the former is invoked only when necessary, i.e., when the lower tier predicts a source.

\vspace{1mm}
\noindent
\textbf{Compression Techniques. }\enspace Sparsity, low-rank, and quantization have been proven to be effective ways of compressing deep architectures like RNNs \cite{wang2017accelerating, ye2018learning} and CNNs \cite{han2015deep_compression, kumari2019edgel}. Many other compression methods like householder reflectors in Spectral-RNN \cite{zhang2018stabilizing}, Kronecker factorization in KRU \cite{jose2017kronecker} have been proposed, which are complementary to the solution proposed in this paper. We incorporate low-rank representation, Q15 quantization, and piecewise-linear approximation \cite{kusupati2018fastgrnn} to make MSC-RNN realizable on Cortex-M3 microcontrollers.

%% file: sysmodel.tex
\section{Radar and Classifier Models}
\label{sec_sysmodel}
\subsection{Micro-power Radar Model}
\label{subsec_radar}

The monostatic PDR sensor depicted in Figure \ref{fig_radar} has a bandwidth of nearly $100$ MHz and a center frequency at about $5.8$ GHz. It is a short-range radar with an anisotropic radiation pattern yielding a maximum detection range of $\tiny{\sim}13$ m. Sensing itself consumes $15$mW of power, not counting the inference computation on the associated microcontroller. The radar response is low pass filtered to 100 Hz; hence the output is typically sampled at rates over 200Hz.

The output signal from the radar is a complex time-series with In-phase (I) and Quadrature (Q) components. When a source moves within the detection range, in addition to the change in received power, the phase of this signal changes according to the direction of motion. Consequently, its relative displacement can be estimated with high accuracy (typically, sub-cm scale), and a rich set of features can be derived by tracking its phase evolution over time.

\subsection{Classifier Architectures}
\subsubsection{Input and Feature Representation}
The radar classifier system uses the aforementioned complex time-series as input. Extant end-to-end architectures for micro-power radar sensing mostly eschew deep feature learning for cheap handcrafted feature engineering in the amplitude, time, and spectral domains \cite{he2014regression, roy2017cross}. However, these solutions incur significant featurization overhead; this is exemplified in Table \ref{tab_latency} on 1-second snippets extracted from the complex time-series. Even ignoring the SVM computation latency, it can be seen that the main computation bottleneck is this incremental overhead which results in >30\% duty cycle on the Cortex-M3, of which $\sim$10\% constitutes the FFT overhead alone. 

\begin{table}[htb]
	\centering
	\vspace*{4mm}
	\caption{Computation overheads in a shallow (SVM) radar solution on Cortex-M3 (10 features, 1s windows)}
	\vspace*{-1.5mm}
	\begin{tabular}{l|r} 
		\hline
		\multicolumn{1}{c|}{\textbf{Component}}    & \multicolumn{1}{c}{\textbf{Latency (ms)}}\\
		\hline\hline
		FFT                 			& 80 \\ 
		Incremental feature computation & 212 \\
		SVM inference	(700 SVs)				& 55 \\
		\hline
	\end{tabular}
	\label{tab_latency}
	\vspace*{-1.5mm}
\end{table}

\subsubsection{Deep Classifier Architecture}

Deep radar classifier systems such as CNNs (or even some RNNs) convert the raw time-series to STFT, and hence also maintain this steady overhead in input representation. In the interest of designing resource efficient solutions, in this work, we instead focus on being competitive with all-domain featurization using purely time-domain learning.

\subsubsection{Shallow Classifier Architecture}
\label{subsec_sysmodel_old}
\begin{figure}[t]
	\centering
	\includegraphics[scale=0.28]{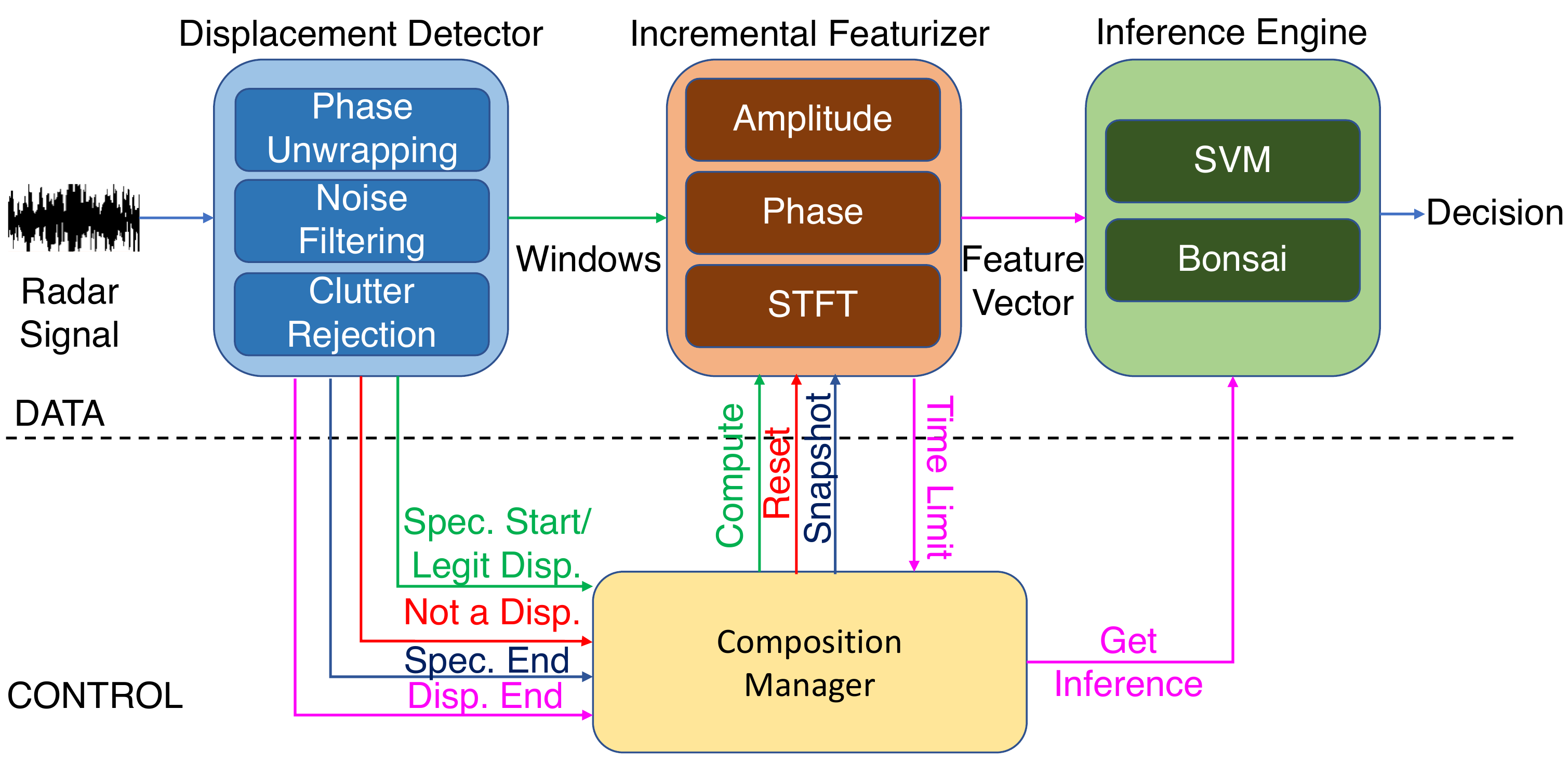}
	\caption{SVM classifier data and control planes; control signal-response pairs are color coded}
	\label{fig_sysmodel_old}
\end{figure}

As shown in Figure \ref{fig_sysmodel_old}, a prototypical shallow radar classifier system consists of three subsystems: (i) a \textit{displacement detector} for discriminating clutter vs.~sources, (ii) an \textit{incremental featurizer}, (iii) an \textit{end inference engine} that discriminates source types, and (iv) a \textit{composition manager} that handles their interactions. The displacement detector is a simple module that thresholds unwrapped phase over incoming windows of radar data ($\frac{1}{2}$ s or 1 s) to detect legitimate source displacements in the scene, filtering in-situ clutter that tends to yield self-canceling phase unwraps. When a source displacement is speculatively detected, the featurizer is invoked till the current displacement ends or a pre-specified time limit is reached. The final feature vector is fed to an end classifier such as Support Vector Machine \cite{roy2017cross}. Note that incremental feature computation overhead \textit{is} the primary impediment in realizing efficiency in these systems, hence techniques like replacing the heavy SVM classifier with the much lighter Bonsai \cite{kumar2017resource}, or observing longer displacements to run inference infrequently do not alleviate this problem.

While we preserve this ontological hierarchy in our solution, we replace this simple ``ensemble" with a principled 2-tier RNN approach in the time-domain. In the next sections, we present our proposed architecture and discuss how deep feature learning can be used to successfully resolve the above issues.

%% file: solarch.tex
\section{2-Tier Deep Classifier Architecture}
\label{sec_solarch}

MSC-RNN is a multi-scale, cascaded architecture that uses EMI-FastGRNN as the lower tier clutter discriminator and FastGRNN as the upper tier source classifier. While EMI-FastGRNN efficiently localizes the source signature in a clutter prone time-series ensuring smaller sequential inputs along with early classification, FastGRNN reduces the per-step computational overhead over much heavier alternatives such as LSTM. We begin with the relevant background for each of these components.

\subsection{Candidate Classifiers}
\label{subsec_components}
\noindent
\textbf{FastGRNN.}\enspace FastRNN \cite{kusupati2018fastgrnn} provably stabilizes RNN training by helping to avoid EVGP by using only two additional scalars over the traditional RNN. FastGRNN is built over FastRNN and it extends the scalars of FastRNNs to vector gates while maximizing the computation reuse. FastGRNN also ensures its parameter matrices are low-rank, sparse and byte quantized to ensure very small models and very fast computation. FastGRNN is shown to match the accuracies of state-of-the-art RNNs (LSTM and GRU) across various tasks like keyword spotting, activity recognition, sentiment analysis, and language modeling while being up to 45x faster.

\begin{figure}[t]
	\centering
	\begin{subfigure}[b]{0.24\textwidth}
		\includegraphics[width=\textwidth]{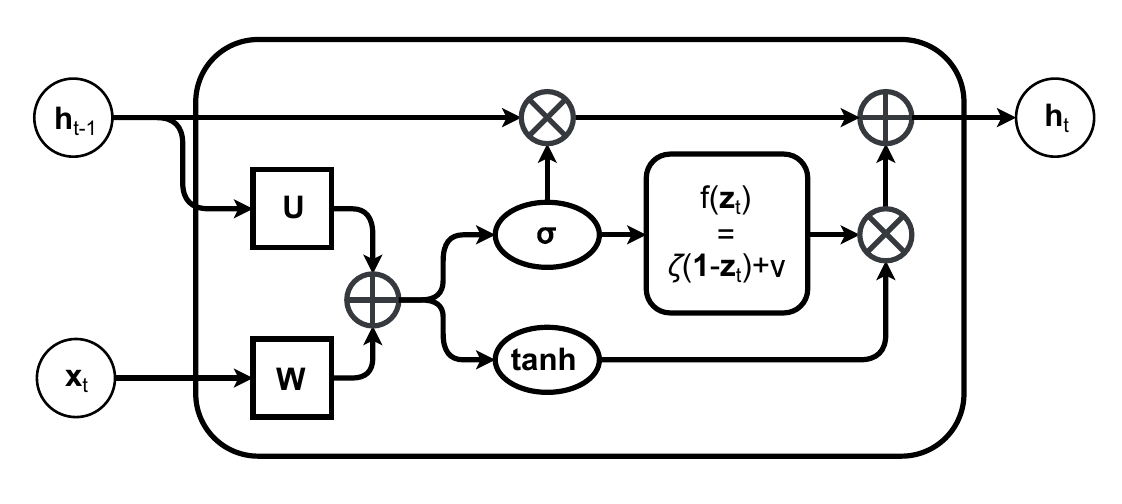}
		\caption{FastGRNN (cascaded)}
	\end{subfigure}%
	~
	\begin{subfigure}[b]{0.23\textwidth}
		\includegraphics[width=\textwidth]{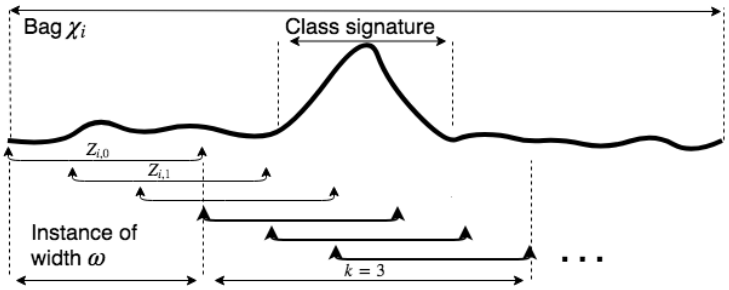}
		\caption{EMI-FastGRNN (always on)}
	\end{subfigure}%
	\caption{FastGRNN \& EMI-FastGRNN (images from \cite{kusupati2018fastgrnn, dennis2018multiple})}
	\label{fig_fastgrnn}
\end{figure}

Let $X=[\vec{x}_1, \vec{x}_2, \hdots, \vec{x}_T]$ be the input time-series, where $x_t \in \mathbb{R}^{\scriptscriptstyle D}$. The traditional RNN's hidden vector $\vec{h}_t \in \mathbb{R}^{\hat{{\scriptscriptstyle D}}}$ captures long-term dependencies of the input sequence: $\vec{h}_t=\tanh(\vec{W}\vec{x}_t+\vec{U}\vec{h}_{t-1}+\vec{b})$. Typically, learning $\vec{U}$ and $\vec{W}$ difficult due to the gradient instability. FastGRNN (Figure \ref{fig_fastgrnn}(a)) uses a scalar controlled peephole connection for every coordinate of $\vec{h}_t$:
\vspace*{-.33mm}
\begin{gather*}
\vec{h}_t = (\zeta(1-\vec{z}_t)+\nu)\odot\tanh(\vec{W}\vec{x}_t+\vec{U}\vec{h}_{t-1}+\vec{b}_h)+\vec{z}_t\odot \vec{h}_{t-1},\\
\vec{z}_t=\sigma(\vec{W}\vec{x}_t+\vec{U}\vec{h}_{t-1}+\vec{b}_z)
\vspace*{-.33mm}
\end{gather*}
Here, $0\leq\zeta,\nu\leq1$ are trainable parameters, and $\odot$ represents the vector Hadamard product. \\

\noindent
\textbf{EMI-RNN.}\enspace Time-series signals when annotated are rarely precise and often coarsely labeled due to various factors like human errors and smaller time frames of activities themselves. EMI-RNN \cite{dennis2018multiple} tackles the problem of signal localization using MIL, by splitting the $i$\textsuperscript{th} data window into instances $\{Z_{i,\tau}\}_{\tau=1,\hdots T-\omega+1}$ of a fixed width $\omega$ (Figure \ref{fig_fastgrnn}(b)). The algorithm alternates between training the classifier and re-labeling the data based on the learned classier until convergence. A simple thresholding scheme is applied to refine the instances: in each iteration, $k$ consecutive instances are found with maximum prediction sum for the class label. Only these instances are included in the training set for the next iteration. Here, $k$ is a hyperparameter that intuitively represents the number of instances expected to cover the source signature. In the end, EMI-RNN produces precise signal signatures which are much smaller than the raw input, thus reducing the computation and memory overhead over the traditional sequential techniques. EMI-RNN also ensures early detection of noise or keywords thereby removing the need of going through the entire signal before making a decision. When combined, EMI-FastGRNN provides very small models along with very fast inference for time-series classification tasks. Codes for FastGRNN \cite{kusupati2018fastgrnn} \& EMI-RNN \cite{dennis2018multiple} are part of Microsoft Research India's EdgeML repository \cite{edgemlcode}.

\subsection{MSC-RNN Design}
\label{subsec_cascade}
While EMI-RNN is by itself equipped to handle multi-class classification efficiently, we find its accuracy and non-dominant source recall to be sub-optimal for the radar time-series, especially at smaller hidden dimensions and shorter window lengths. FastGRNN, on the other hand, is a relatively heavier solution to be used as a continuously running 3-class discriminator. To redress this trade-off, we make the following observations:

\begin{enumerate}[label=(\roman*)]
	\item clutter, which yields self-canceling phase, can be rejected at a relatively shorter time-scale,
	\item disambiguating source types from their complex returns is a harder problem requiring a potentially longer window of observation, and
	\item the common case in a realistic deployment constitutes clutter; legitimate displacements are relatively few.
\end{enumerate}

\begin{figure}[t]
	\centering
	\includegraphics[scale=0.3]{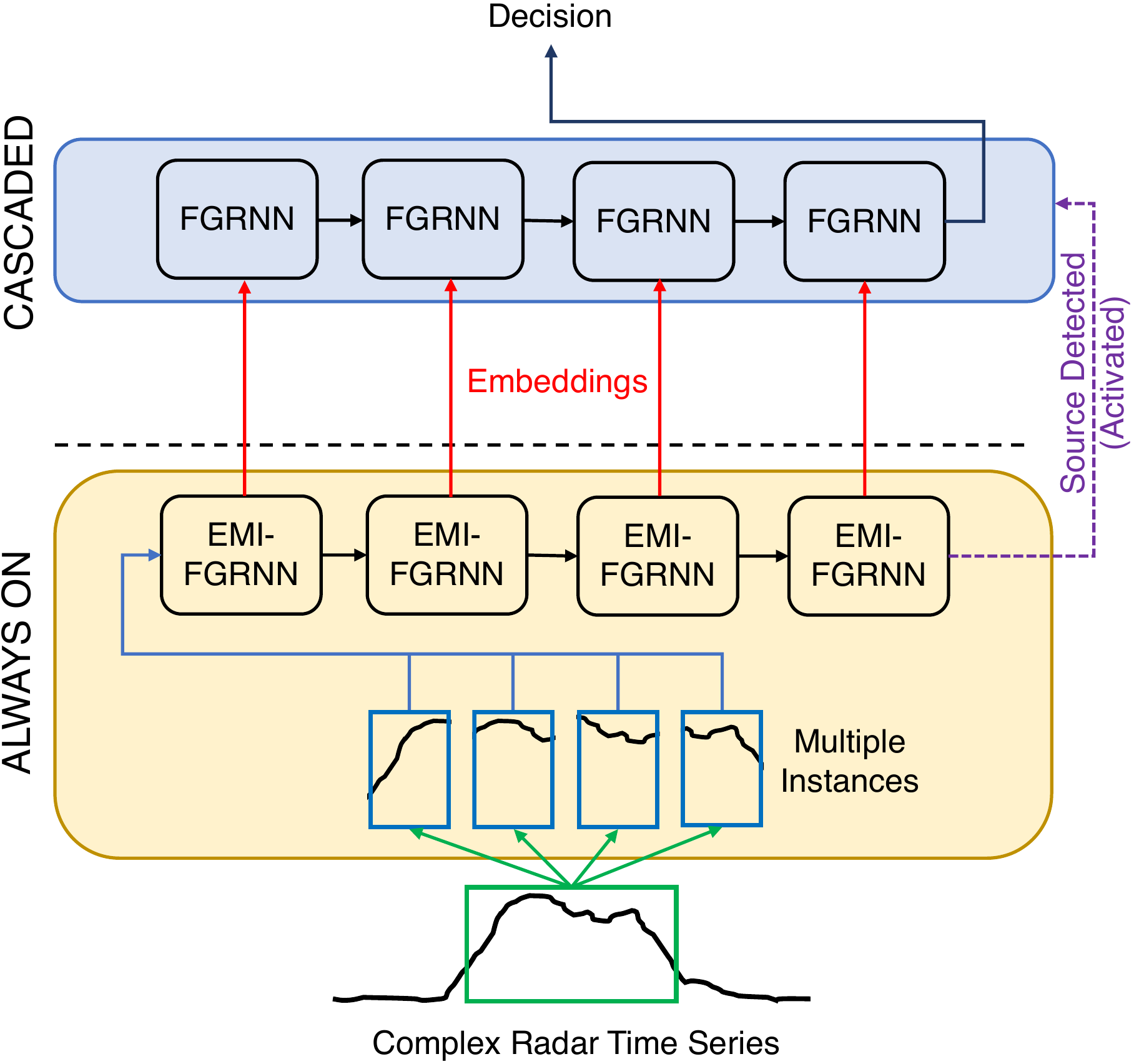}
	\caption{MSC-RNN architecture -- the lower EMI-FastGRNN runs continuously, while the higher FastGRNN is invoked only for legitimate displacements}
	\label{fig_solarch}
\end{figure}

MSC-RNN, therefore, handles the two sub-problems at different time-scales of featurization (see Figure \ref{fig_solarch}): the lower tier, an EMI-FastGRNN, discriminates sources from  clutter at the level of short instances, while the upper one, a windowed FastGRNN, discriminates source types at the level of longer windows. Further, the upper tier is invoked \textit{only} when a source is discriminated by the lower tier and operates on the instance-level embeddings generated by the latter. 

\subsubsection{Joint Training and Inference}
The training of the lower tier inherits from that of EMI-training. We recap its training algorithm \cite{dennis2018multiple}, which occurs in two phases, the MI phase and the EMI phase. In the MI phase, where the source boundaries are refined in a clutter-prone window, the following objective function is optimized:
$$\min\limits_{f_l,s_i}\frac{1}{n}\sum\limits_{i,\tau}\mathbbm{1}_{\tau\in[s_i, s_i+k]}\ell(f_l(Z_{i,\tau}),y_i)$$

Here, $\ell$ represents the loss function of FastGRNN, and the classifier $f_l$ is based on the final time-step in an instance. In the EMI phase, which incorporates the early stopping, the loss $\mathcal{L}_\text{EMI}$ is obtained by replacing the previous loss function with the sum of the classifier loss at every step: $\min\sum\limits_i\sum\limits_{t=1}^T\ell(w^To_{i,t})$, where $w$ is the fully connected layer and $o_{i,t}$ the output at step $t$. The overall training proceeds in several rounds, where the switch to the EMI loss function is typically made halfway in. 

\begin{figure*}[t]
	\begin{subfigure}[b]{0.21\textwidth}
		\centering
		\includegraphics[width=\textwidth]{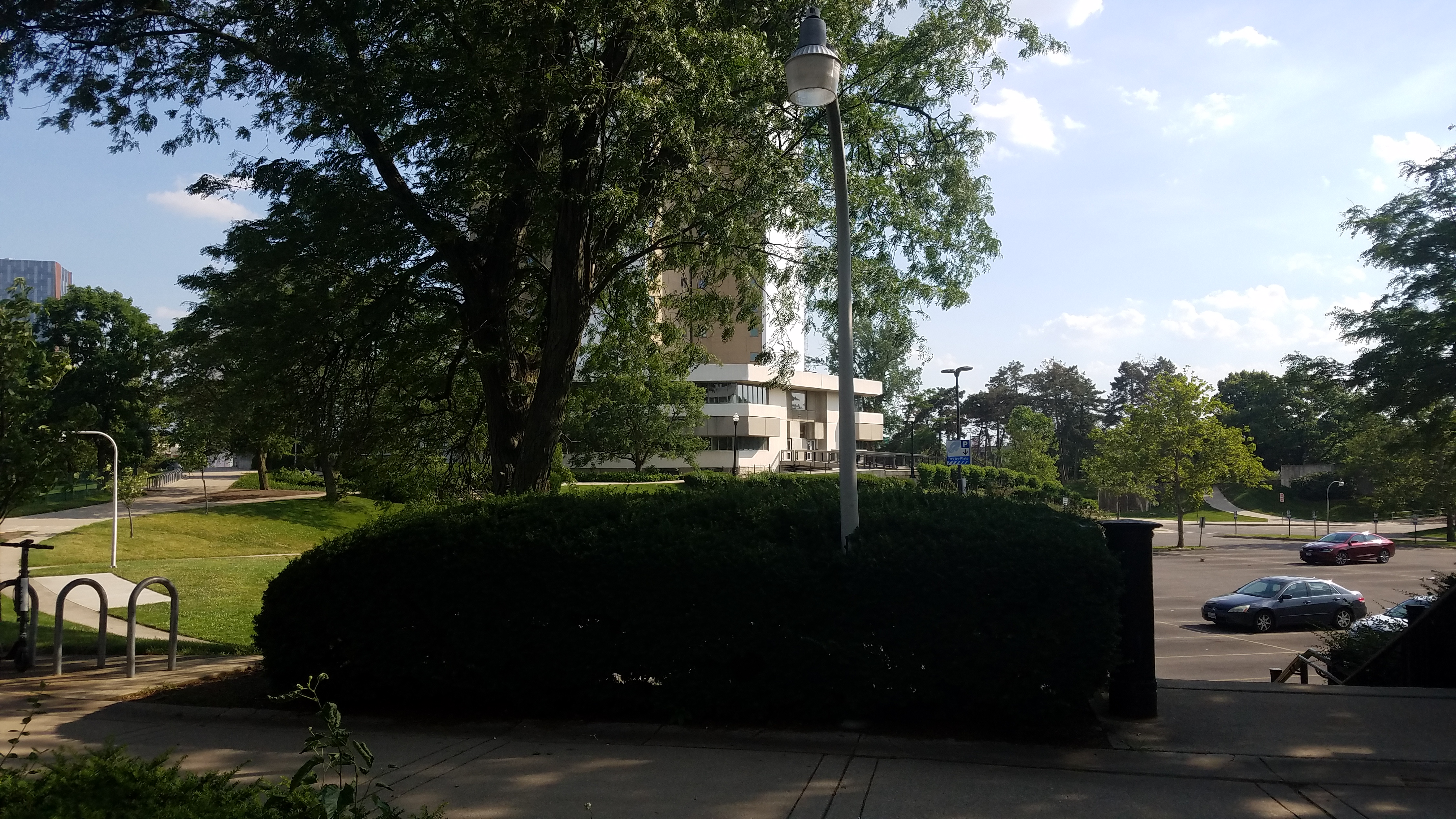}
		\caption{Public park}
	\end{subfigure}%
	~
	\begin{subfigure}[b]{0.21\textwidth}
		\centering
		\includegraphics[width=\textwidth]{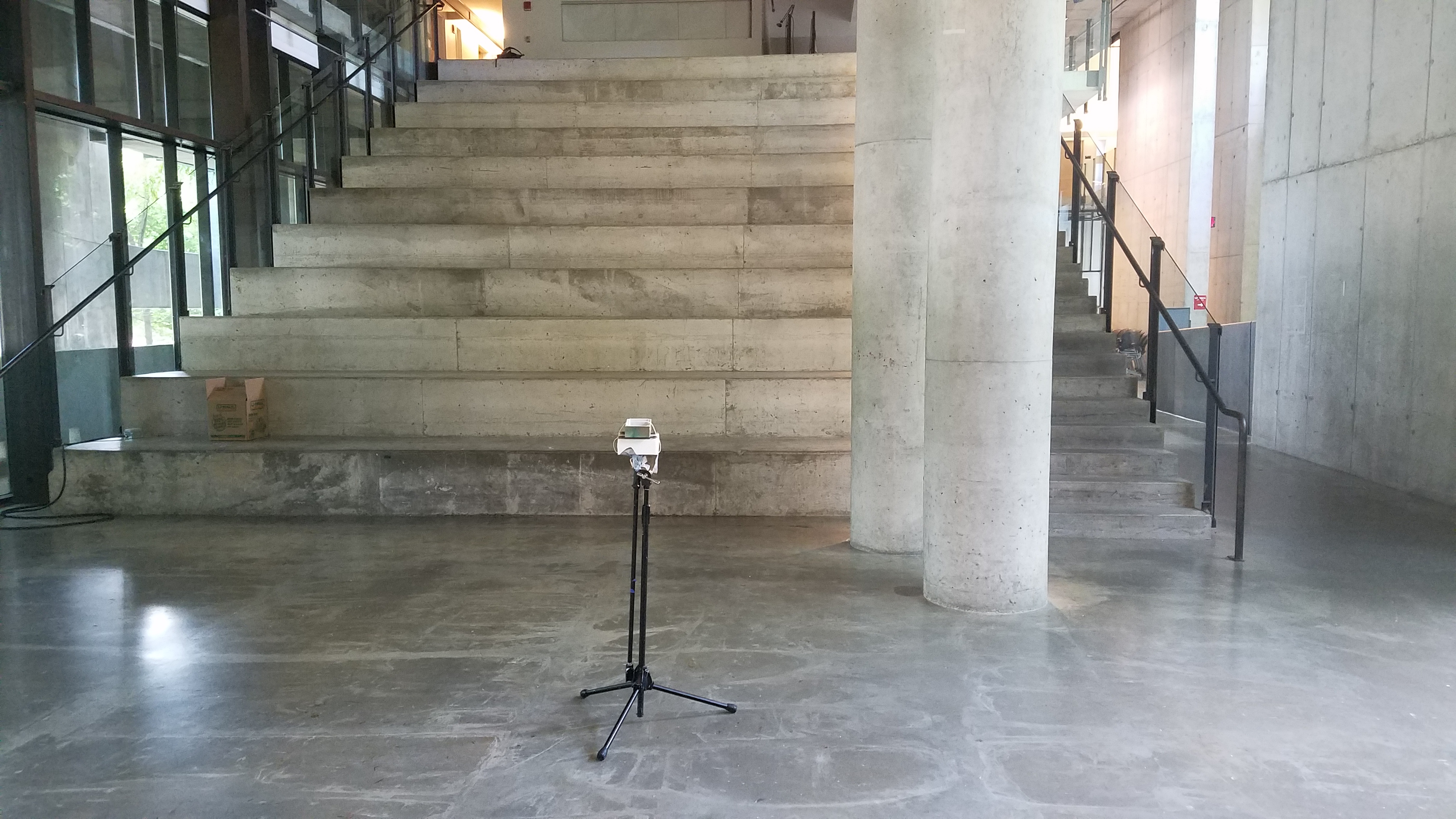}
		\caption{Indoor amphitheater}
	\end{subfigure}
	~
	\begin{subfigure}[b]{0.21\textwidth}
		\centering
		\includegraphics[width=\textwidth]{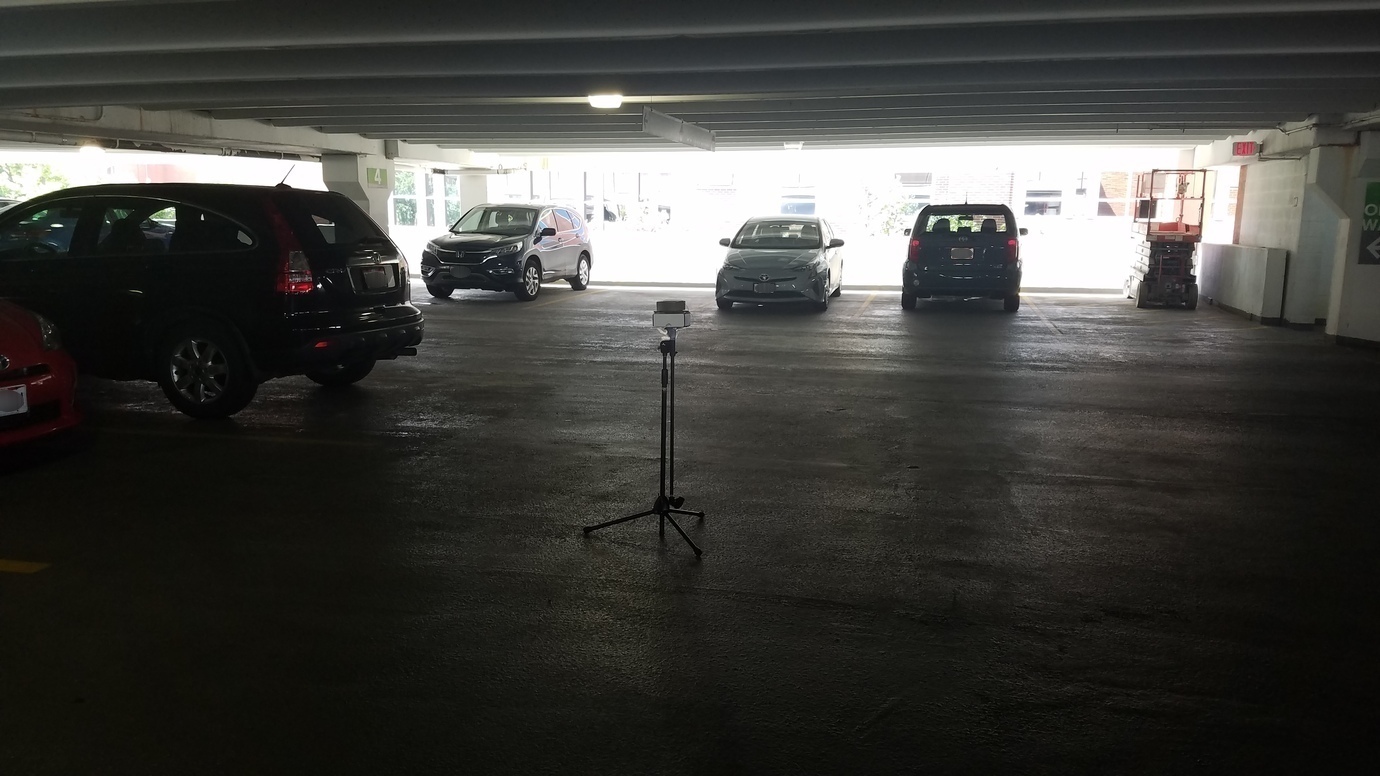}
		\caption{Parking garage bldg.}
	\end{subfigure}
	~
	\begin{subfigure}[b]{0.21\textwidth}
		\centering
		\includegraphics[width=\textwidth]{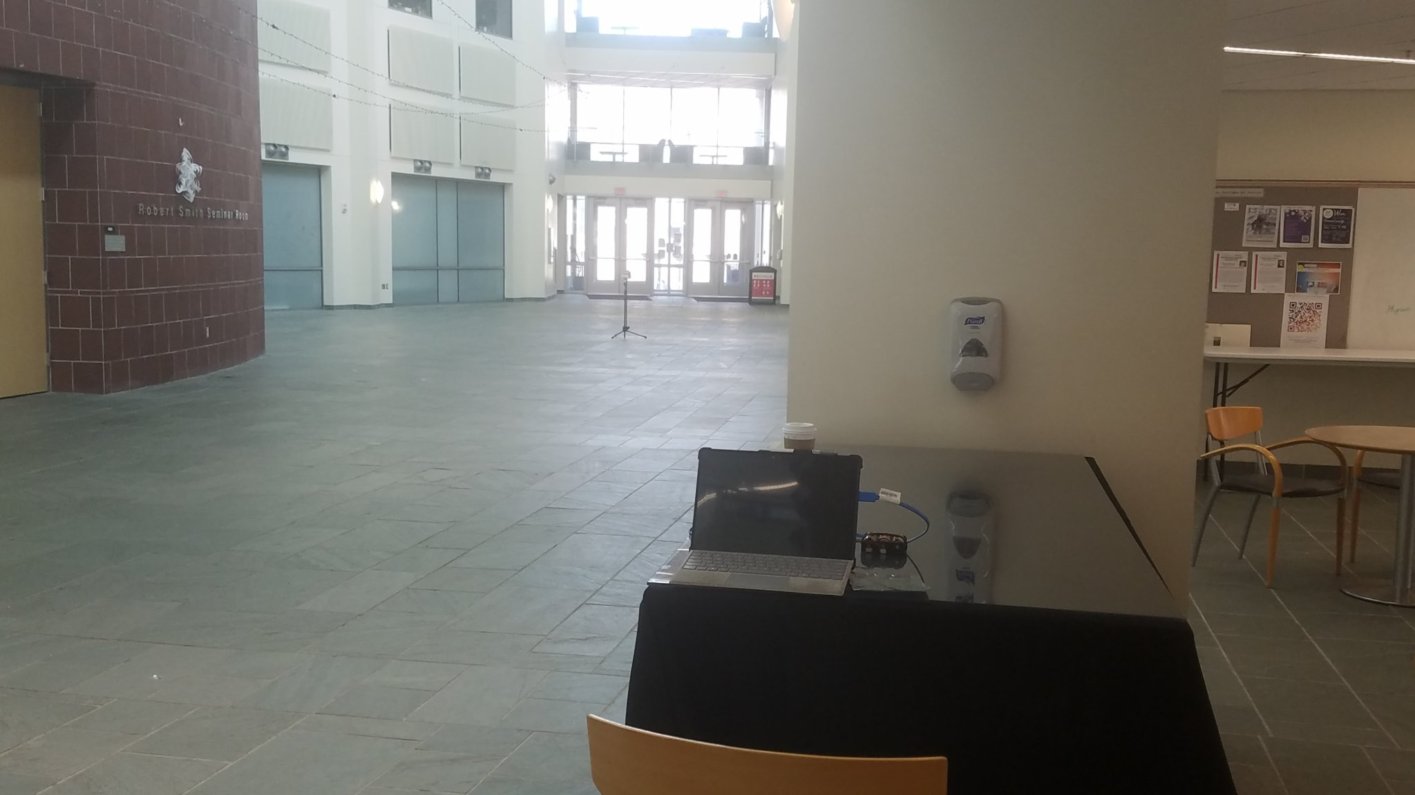}
		\caption{Building foyer}
	\end{subfigure}
	\caption{Some locations where source and clutter data was collected for experiments}
	\label{fig_locations}
\end{figure*}

For training the upper tier, in keeping with the divide-and-conquer paradigm of MSC-RNN, the upper tier FastGRNN cell should only learn to separate the source types, while ignoring instances of training data that are clutter. Therefore, we devise a conditional training strategy that captures the cascading behavior. To achieve this, the standard cross-entropy loss function of the upper tier is modified as:
$$\min\limits_{f_u}\frac{1}{n}\sum\limits_{i}\mathbbm{1}_{y_i\neq -1}\ell(f_u(\mathcal{E}(\{Z^{tr}_{i,\tau}\})),y_i)$$
where $f_u$ represents the upper classifier, and $\mathcal{E}:\mathbb{R}^{(T-\omega+1)\times \omega\times F}\rightarrow\mathbb{R}^{(T-\omega+1)\times H_l}$ represents the instance-level embedding vector from EMI-RNN with a hidden dimension of $H_l$ (here, $F$ represents the feature dimension for the radar time-series). Intuitively, this means that the upper loss is unaffected by clutter points, and thus the tiers can be kept separate.

\begin{algorithm}[htb]
	\caption{MSC-RNN training algorithm}
	\begin{algorithmic}
		\REQUIRE Multi-instance training data $\{\{Z^{tr}_{i,\tau}\}_{\tau=1,\hdots T-\omega+1}, y^{tr}_i\}_{i=1,\hdots n}$, the number of rounds $n_r$, $k$\\
		
		\renewcommand{\algorithmicensure}{\textbf{Training:}}
		\ENSURE
		\STATE Freeze FastGRNN, unfreeze EMI-FastGRNN\\
		\REPEAT
		\STATE Train \textbf{EMI-FastGRNN}($\{\{Z^{tr}_{i,\tau}\}, \mathbbm{1}_{y^{tr}_i\neq -1}\}$)
		\UNTIL{convergence}
		
		\STATE Freeze EMI-FastGRNN, unfreeze FastGRNN\\
		\REPEAT
		\STATE Train \textbf{FastGRNN}($\{\mathcal{E}(\{Z^{tr}_{i,\tau}\},y^{tr}_i)\}$), minimizing loss $\frac{1}{n}\sum\limits_{i}\mathbbm{1}_{y_i\neq -1}\ell(f_u(\mathcal{E}(\{Z^{tr}_{i,\tau}\})),y_i)$
		\UNTIL{convergence}
		
		\STATE Unfreeze both EMI-FastGRNN and FastGRNN
		
		\FOR{$r\in n_r$}
		\IF{$r<\frac{n_r}{2}$}
		\STATE $\mathcal{L}_\text{lower}\leftarrow$ MI-loss
		\ELSE
		\STATE $\mathcal{L}_\text{lower}\leftarrow$ EMI-loss
		\ENDIF
		\REPEAT
		\STATE Train \textbf{MSC-RNN}($\{\{Z^{tr}_{i,\tau}\}, \mathbbm{1}_{y^{tr}_i\neq -1}\}$) minimizing loss $\mathcal{L}_\text{lower}+\frac{1}{n}\sum\limits_{i}\mathbbm{1}_{y_i\neq -1}\ell(f_u(\mathcal{E}(\{Z^{tr}_{i,\tau}\})),y_i)$
		\UNTIL{convergence}
		\ENDFOR
	\end{algorithmic}
	\label{alg_cascade}
\end{algorithm}

The training algorithm for MSC-RNN is outlined in Algorithm \ref{alg_cascade}. The two tiers are first separately initialized using their respective loss functions, and in the final phase, both are jointly trained to minimize the sum of their losses. Inference is simple: the instance-level EMI-RNN stops early with a decision of ``Source'' when a probability threshold $\hat{p}$ is crossed; $\geq k$ consecutive positives constitute a discrimination for which the cascade is activated.


%% file: evaluation.tex
\section{Comparative \& Tier-wise Evaluation}
\label{sec_eval}
\subsection{Datasets}
\begin{figure}[b]
	\begin{subfigure}[t]{0.22\textwidth}
		\centering
		\includegraphics[width=\textwidth]{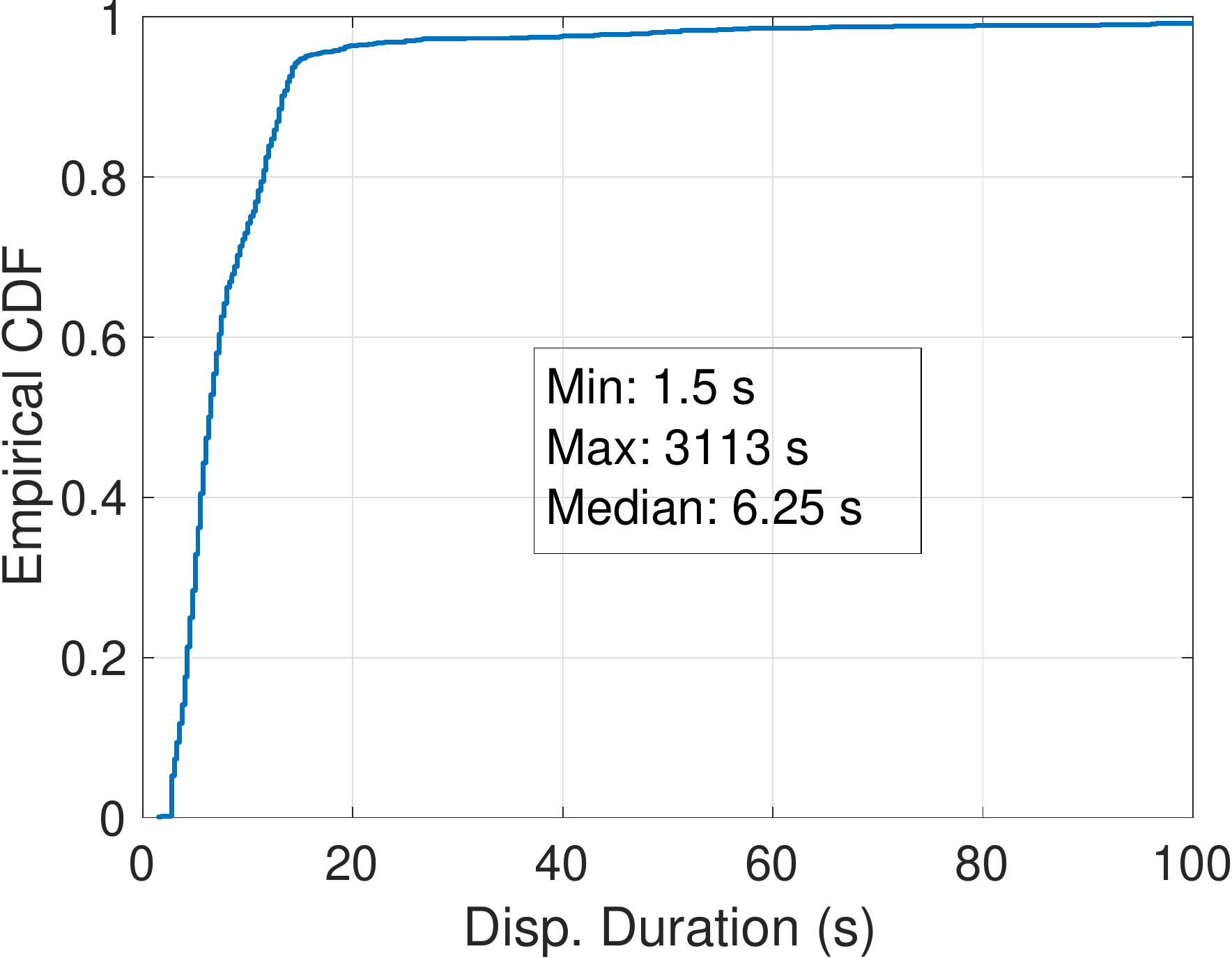}
		\caption{Disp. Duration CDF (partial)}
	\end{subfigure}%
	~
	\begin{subfigure}[t]{0.225\textwidth}
		\centering
		\includegraphics[width=\textwidth]{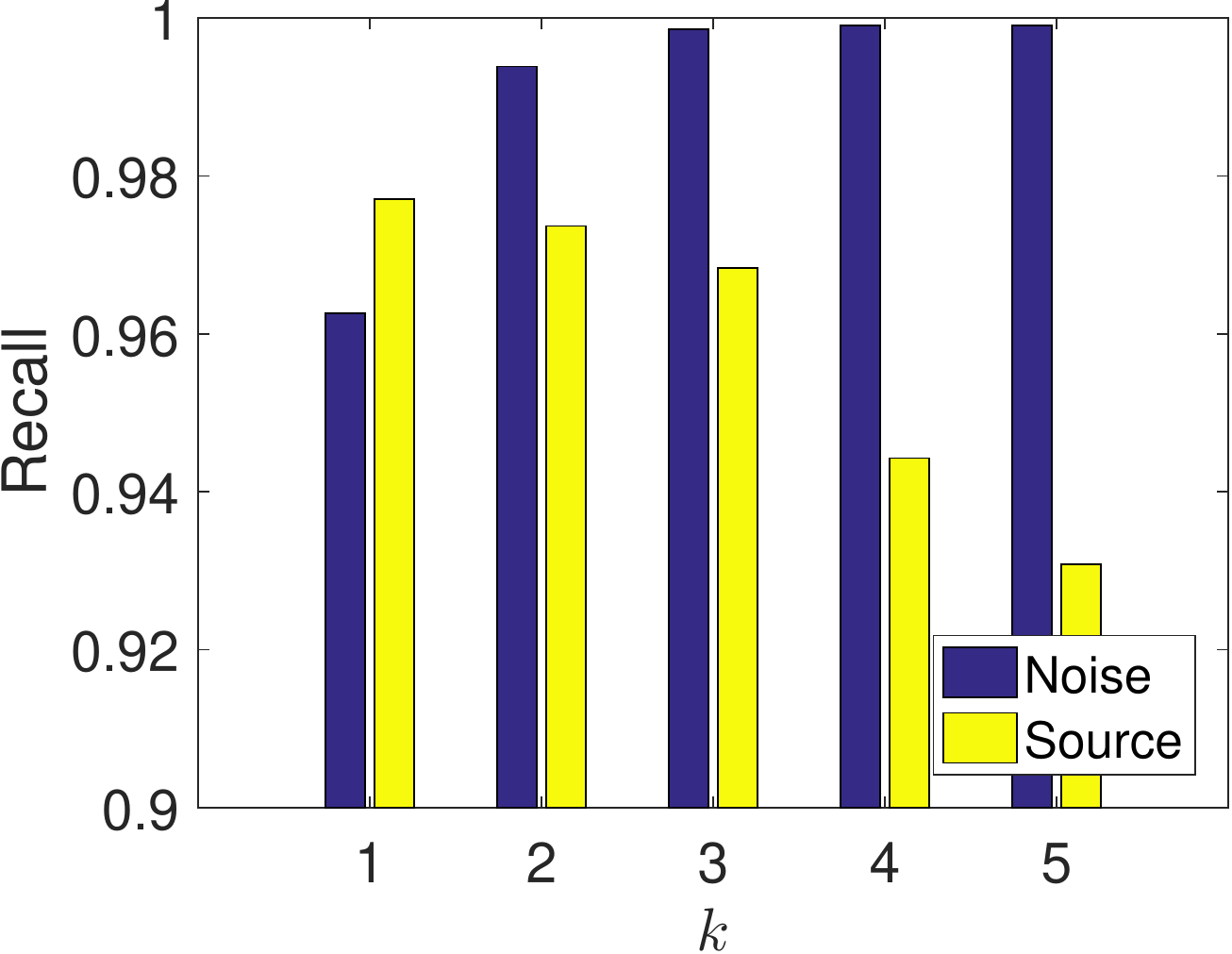}
		\caption{Impact of $k$ on EMI Recalls}
	\end{subfigure}
	\caption{Source detected duration CDF for the data in Table \ref{tab_datasets_cuts} and how the hyperparameter $k$ in 2-class EMI affects their detection (1 second windows)}
	\label{fig_design_choices}
\end{figure}

Table \ref{tab_datasets_cuts} lists the radar source and clutter datasets collected in various indoor and outdoor environments, which are used in this work.
Some of these locations are documented in Figure \ref{fig_locations}; small or crammed indoor spaces such as office cubicles have been avoided to prevent the radar returns from being adversely affected by multi-path effects and because they are not central to the smart city scenarios. A partial distribution of displacement durations is provided in Figure \ref{fig_design_choices}(a).  Each data collect has associated with it the corresponding ground truth, recorded with motion-activated trail cameras or cellphone video cameras, with which the radar data was correlated offline to ``cut'' and label the source displacement snippets appropriately\footnote{The radar dataset, which we have open-sourced, does not include individually identifiable information of living individuals and is thus not considered research with human subjects per 45 CFR $\S$46.102(e)(1)(ii).}. The datasets have been balanced in the number of human and non-human displacement points where possible, and windowed into snippets of 1, 1.5, and 2 seconds which correspond to 256, 384, and 512 I-Q sample pairs respectively. We note that due to the duration of collections and differences in average displacement lengths, etc., humans are underrepresented in these datasets compared to the other labels.
Table \ref{tab_datasets_windows} shows the number of training, validation, and test points for each of these window lengths on a roughly 3:1:1 split. Currently, only the cattle set has multiple concurrent targets; efforts to expand our datasets with target as well as radar type variations are ongoing.

\begin{table}[htb]
	\caption{Radar evaluation datasets}
	\begin{subtable}{0.45\textwidth}
	\centering
	\caption{Source displacement counts and clutter durations}
	\begin{tabular}{l|ll} 
		\hline
		\multirow{2}{*}{\textbf{Env.}} & \multicolumn{2}{c}{\textbf{Data Type}}  \\ 
		\cline{2-3}
		& \textbf{Type}   & \textbf{Count}        \\ 
		\hline\hline
		Building foyer                 & Human, Gym ball & 52, 51                \\
		Indoor amphitheater            & Human, Gym ball & 49, 41               \\
		Parking garage bldg.           & Human           & 268                   \\
		Parking lot                    & Human, Car      & 50, 41                \\
		Indoor soccer field            & Human, Gym ball & 90, 82                \\
		Large classroom                & Human, Gym ball & 48, 50                \\
		Cornfield                      & Human, Dog      & 117, 85               \\
		Cattle shed                    & Cow             & 319                   \\ 
		\hline
		Playground		               & Clutter           & 45 mins               \\
		Parking garage bldg.           & Clutter           & 45 mins               \\
		Public park              & Clutter           & 45 mins               \\
		Garden            & Clutter           & 45 mins               \\
		Lawn                          & Clutter           & 20 mins               \\
		\hline
	\end{tabular}
	\label{tab_datasets_cuts}
	\end{subtable}%
	\vspace{5mm}
	\begin{subtable}{0.45\textwidth}
	\centering
	\caption{Windowed data from (a) showing number of training, validation, and test points}
	\begin{tabular}{l|lll} 
		\hline
		\multirow{2}{*}{\textbf{Window Len. (s)}} & \multicolumn{3}{c}{\textbf{\#Windows}}                      \\ 
		\cline{2-4}
		& \textbf{Training} & \textbf{Validation} & \textbf{Testing}  \\ 
		\hline\hline
		1                                        & 17055             & 5685                & 5685              \\
		1.5                                      & 11217             & 3739                & 3739              \\
		2                                        & 8318              & 2773                & 2773              \\
		\hline
	\end{tabular}
	\label{tab_datasets_windows}
	\end{subtable}
\label{tab_datasets}
\end{table}

\subsection{Evaluation Methodology}
\label{subsec_eval_method}
Our proposed architecture is compared with existing shallow radar solutions that use feature handcrafting in the amplitude, phase and spectral domains, as well as with other MIL RNNs. In all cases involving RNNs, the radar data is represented purely in the time-domain. The models chosen for this evaluation are:

\begin{enumerate}[label=(\alph*)]
	\item \textbf{2-tier SVM with phase unwrapped displacement detection.}\enspace Phase unwrapping \cite{goldstein1988satellite} is a widely used technique in radar displacement detection due to its computational efficiency. The idea is to construct the relative trajectory of a source by accumulating differences in successive phase measurements, whereby clutter can be filtered out. We contrast MSC-RNN with a two-tier solution proposed in \cite{roy2017cross}, which uses a robust variant of phase unwrapping with adaptive filtering of clutter samples.
	
	\item \textbf{3-class SVM.}\enspace A clutter vs human vs non-human SVM solution that uses feature handcrafting.
	
	\item \textbf{EMI-FastGRNN.}~An EMI version of FastGRNN (Section \ref{sec_solarch}).
	
	\item \textbf{EMI-LSTM.}~An EMI version of the LSTM. Note that this is a much heavier architecture than the former, and should not be regarded as suitable for a microcontroller device.
\end{enumerate}

Since shallow featurization incurs high incremental overhead, real-time micro-power radar solutions typically avoid techniques such as PCA \cite{candes2011robust}, logistic regression \cite{james2013introduction} or low-dimensional projection \cite{kumar2017resource}. Instead, the 15 best features are selected using the \textit{Max. Relevance, Min. Redundancy} (mRMR) \cite{peng2005feature} algorithm.

For the MIL experiments, the windowed data from Table \ref{tab_datasets_windows} is further reshaped into instances of length 48$\times$2 samples with a fixed stride of 16$\times$2, where 2 refers to the number of features (I and Q components of radar data). For example, for 1 second windows, the shape of the training data for MIL experiments is (17055, 14, 48, 2), and the shape of the corresponding instance-level one-hot labels is (17055, 14, 3). In the interest of fairness and also to avoid a combinatorial exploration of architectural parameters, we present results at fixed hidden sizes of 16, 32, and 64. For MSC-RNN, the lower tier's output (embedding) dimension and upper tier's hidden dimension are kept equal; however, in practice, it is easy to parameterize them differently since the former only affects the latter's input dimension.

\subsubsection{Hyperparameters}
Table \ref{tab_hyperparams} lists the hyperparameter combinations used in our experiments. For the upper-tier source discrimination comparison in Section \ref{subsec_eval_components}, FastGRNN is also allowed to select its optimum input length from 16, 32, and 64 samples.

The selection of the EMI hyperparameter $k$ merits some discussion, in that it controls the extent of ``strictness" we assign to the definition of \emph{displacement}. A higher $k$ makes it more difficult for a current window to be classified as a source unless the feature of interest is genuinely compelling. Expectedly, this gives a trade-off between clutter and source recall as is illustrated in Figure \ref{fig_design_choices}(b). As explained in Section \ref{sec_intro}, controlling for false positives is extremely important in radar sensing contexts such as intrusion detection. Hence, we empirically set $k$ to 10, the smallest value that gives a clutter recall of 0.999 or higher in our windowed datasets.

\begin{table}[htb]
	\caption{Training hyperparameters used}
	\begin{tabular}{l|l|l}
		\hline
		\multicolumn{1}{c|}{\textbf{Model}} & \multicolumn{1}{c|}{\textbf{Hyperparameter}} & \multicolumn{1}{c}{\textbf{Values}}                                                    \\ \hline\hline
		\multirow{5}{*}{EMI/FastGRNN}       & Batch Size                                   & 64, 128                                                                                \\
		& Hidden Size                                  & 16, 32, 64                                                                             \\
		& Gate Nonlinearity                                      & sigmoid, tanh                                                                          \\
		& Update Nonlinearity                                    & sigmoid, tanh                                                                          \\
		& $k$											& 10\\
		& Keep prob. (EMI-LSTM)                                    & 0.5, 0.75, 1.0 \\
		& Optimizer                                    & Adam                                                                                   \\ \hline
		\multirow{2}{*}{SVM}                & $c$                                            & \begin{tabular}[c]{@{}l@{}}0.001,0.01,0.1,1,\\ 10,100,1e3,1e5,1e6\end{tabular} \\
		& $\gamma$                                     & \begin{tabular}[c]{@{}l@{}}0.001,0.01,0.05,\\ 0.1,0.5,1,5,10\end{tabular}                                                     \\ \hline
	\end{tabular}
	\label{tab_hyperparams}
\end{table}

\subsection{Results}
\subsubsection{Comparative Classifier Performance}
\label{subsec_svm_vs_msc}
We compare the inference accuracy and recalls of MSC-RNN, with the RNN and shallow solutions outlined in Section \ref{subsec_eval_method}.

\begin{figure}[htb]
	\begin{subfigure}[t]{0.23\textwidth}
		\centering
		\includegraphics[width=\textwidth]{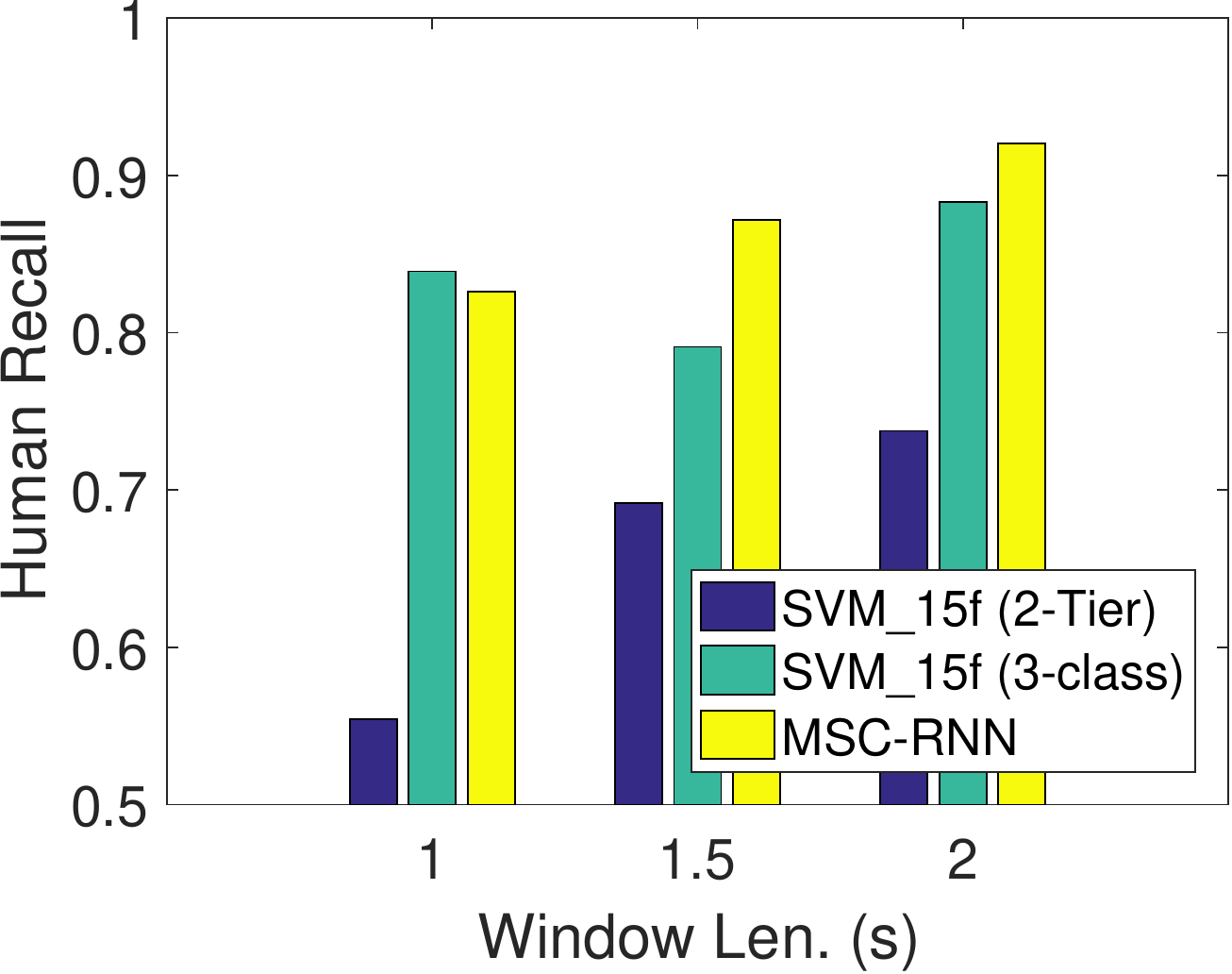}
		\caption{Human Recall}
	\end{subfigure}
	~
	\begin{subfigure}[t]{0.23\textwidth}
		\centering
		\includegraphics[width=\textwidth]{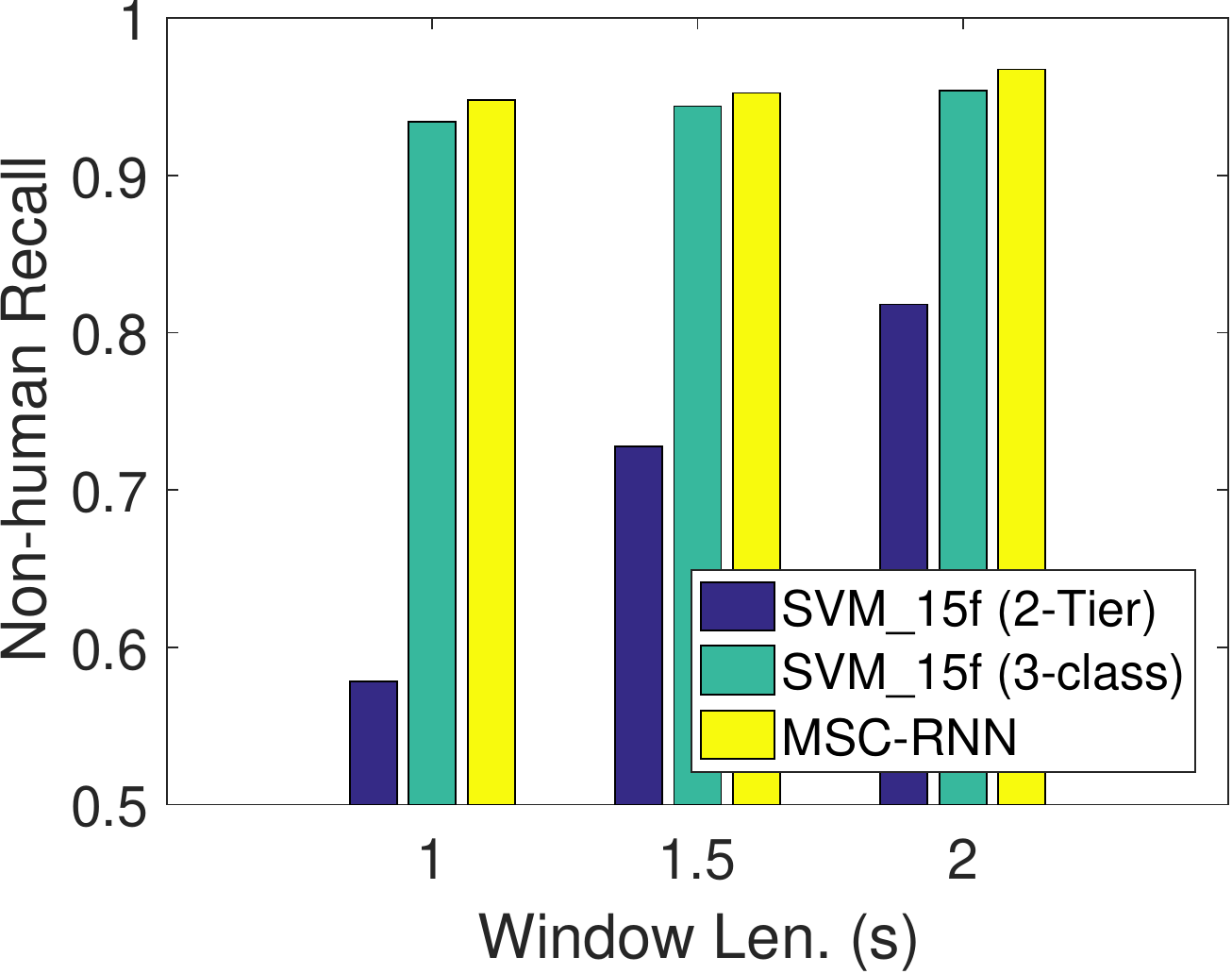}
		\caption{Non-human Recall}
	\end{subfigure}
	
	\vspace{5mm}
	\begin{subfigure}[t]{0.5\textwidth}
		\centering
		\begin{tabular}{l|cc|cc}
			\hline
			\multicolumn{1}{c|}{\multirow{2}{*}{\textbf{\begin{tabular}[c]{@{}c@{}}Win.\\ Len. (s)\end{tabular}}}} & \multicolumn{2}{c|}{\textbf{Accuracy}}                                                                                                                                                & \multicolumn{2}{c}{\textbf{Clutter Recall}}                                                                                                                                            \\ \cline{2-5} 
			\multicolumn{1}{c|}{}                                                                                  & \multicolumn{1}{c}{\textbf{\begin{tabular}[c]{@{}c@{}}SVM\_15f\\ (3-class)\end{tabular}}} & \multicolumn{1}{c|}{\textbf{\begin{tabular}[c]{@{}c@{}}MSC-RNN \end{tabular}}} & \multicolumn{1}{c}{\textbf{\begin{tabular}[c]{@{}c@{}}SVM\_15f\\ (3-class)\end{tabular}}} & \multicolumn{1}{c}{\textbf{\begin{tabular}[c]{@{}c@{}}MSC-RNN\\ \end{tabular}}} \\ \hline\hline
			1    & 0.851   & 0.944   & 0.758   & 0.999   \\
			1.5  & 0.934   & 0.954   & 0.996   & 0.999   \\
			2    & 0.959   & 0.972   & 0.999   & 1.000   \\ \hline
		\end{tabular}
		\caption{Accuracy and Clutter Recall (3-class SVM and MSC-RNN)}
	\end{subfigure}
	\caption{Classification comparison of purely time-domain FastGRNN with two SVM solutions: (a) a 2-tier system using a phase unwrapped clutter rejector as the lower tier, and (b) a 3-class SVM. Both use 15 high information features handcrafted in the amplitude, time, and spectral domains}
	\label{fig_shallow_vs_deep}
\end{figure}

Recall that we have purposefully devised a purely time-domain solution for source discrimination for efficiency reasons, since one of the main components of featurization overhead is that of FFT computations. Figure \ref{fig_shallow_vs_deep} compares MSC-RNN with engineered features in the amplitude, time, and spectral domains that are optimized for micro-power radar classification. For the two-tier SVM, the source recalls for increasing window sizes are inferred from Figure \ref{fig_missed_prob} (discussed in Section \ref{subsec_eval_components}). We find that MSC-RNN significantly outperforms the 2-tier SVM solution in terms of human and non-human recalls, even with features learned from the raw time-series. Similarly, for the 3-class case, our solution provides much more stable noise robustness and is generally superior even to the much heavier SVM solution.

\begin{figure*}[t]
	\begin{subfigure}[h]{0.25\textwidth}
		\centering
		\includegraphics[width=\textwidth]{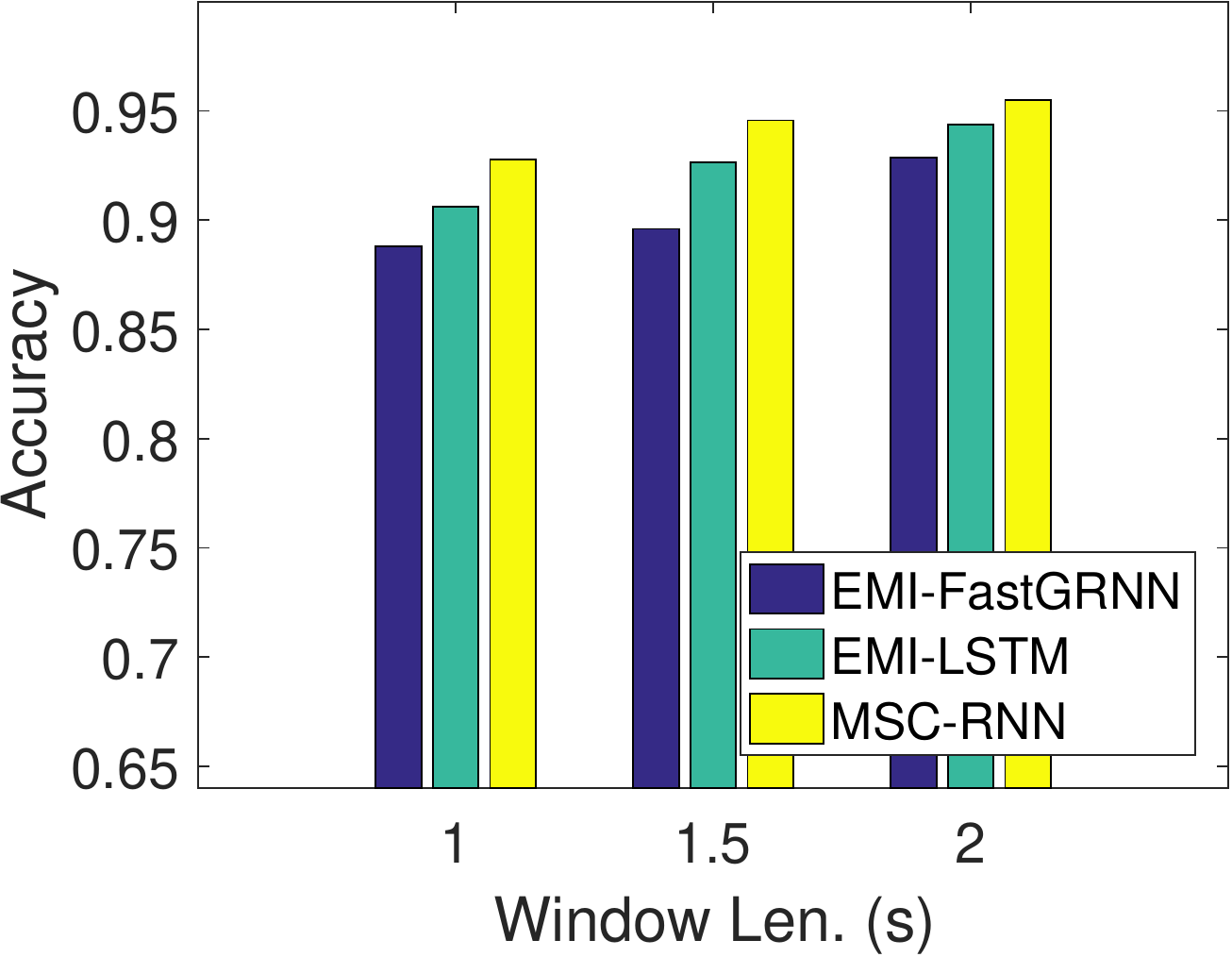}
		\caption{Accuracy (H=16)}
	\end{subfigure}%
	~
	\begin{subfigure}[h]{0.25\textwidth}
		\centering
		\includegraphics[width=\textwidth]{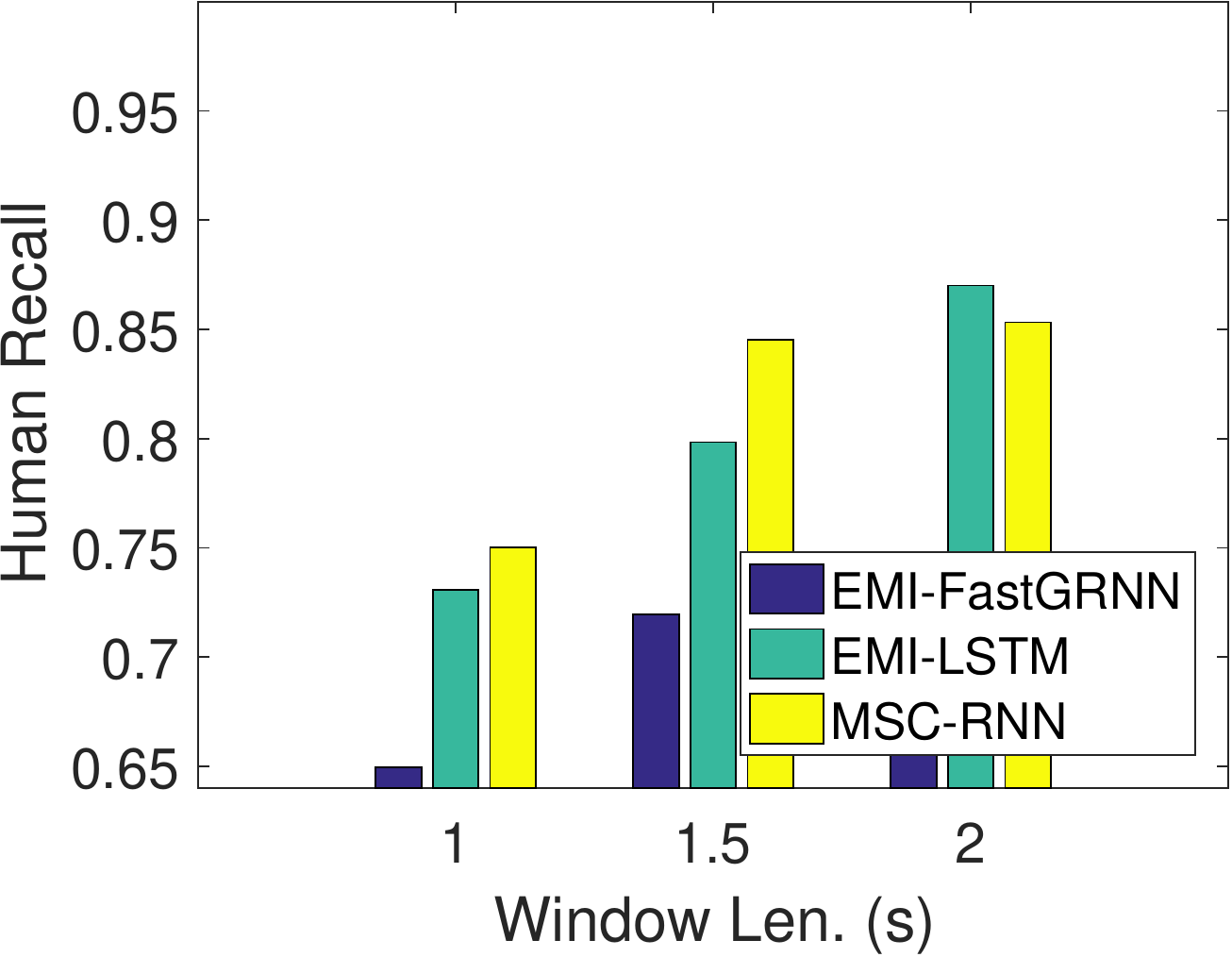}
		\caption{Human Recall (H=16)}
	\end{subfigure}%
	~
	\begin{subfigure}[h]{0.25\textwidth}
		\centering
		\includegraphics[width=\textwidth]{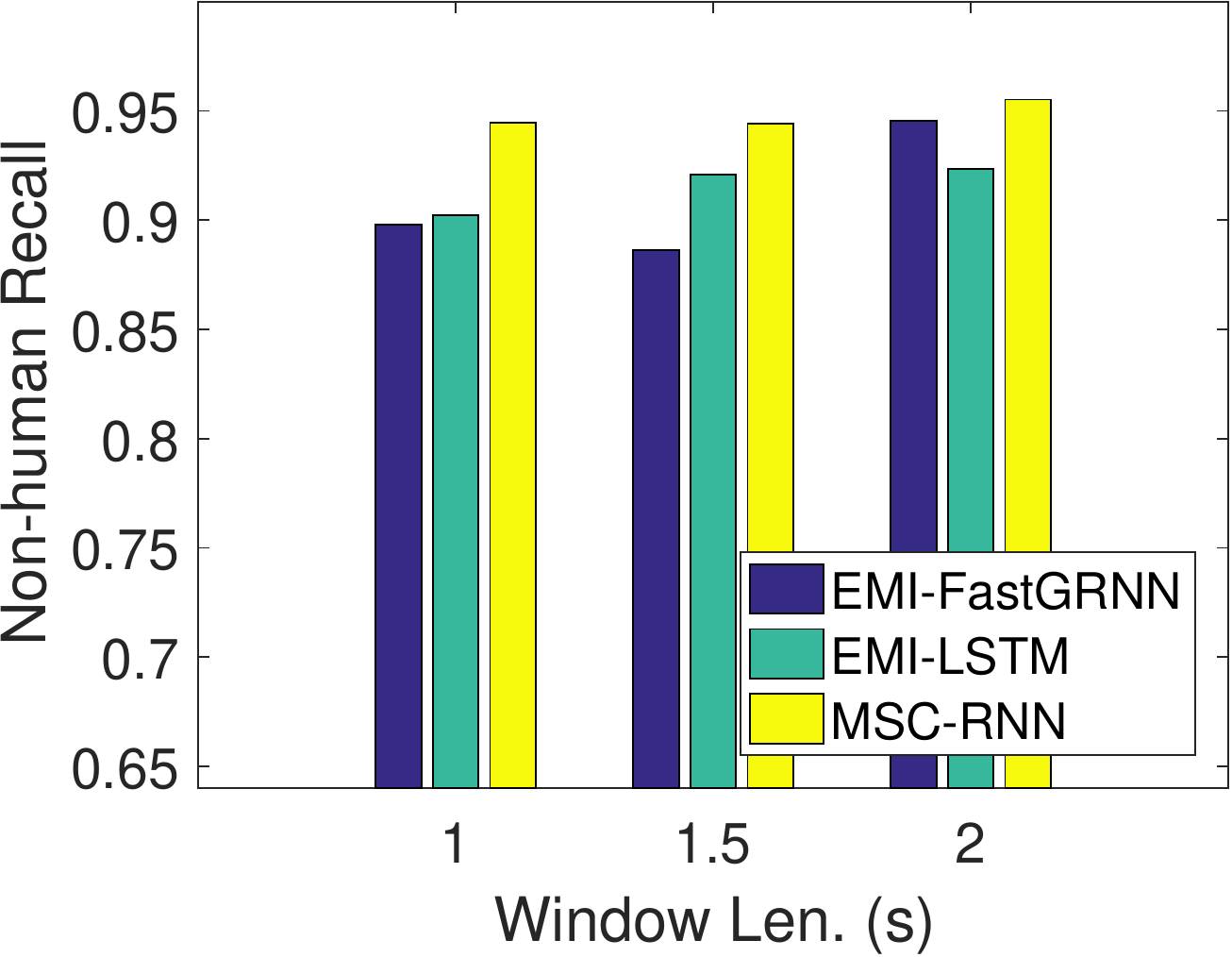}
		\caption{Non-human Recall (H=16)}
	\end{subfigure}%
	
	\vspace{5mm}
	\begin{subfigure}[h]{0.25\textwidth}
		\centering
		\includegraphics[width=\textwidth]{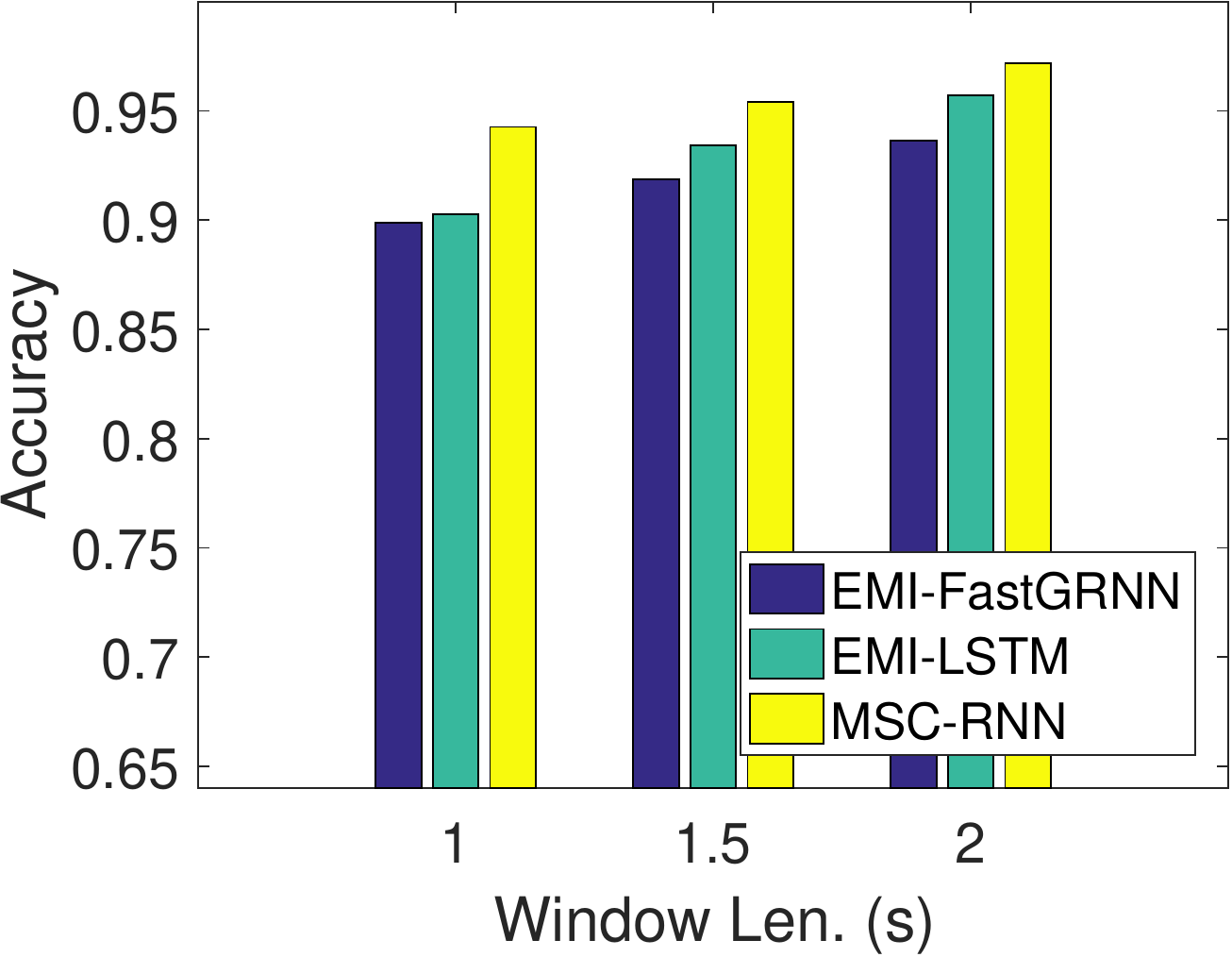}
		\caption{Accuracy (H=32)}
	\end{subfigure}%
	~
	\begin{subfigure}[h]{0.25\textwidth}
		\centering
		\includegraphics[width=\textwidth]{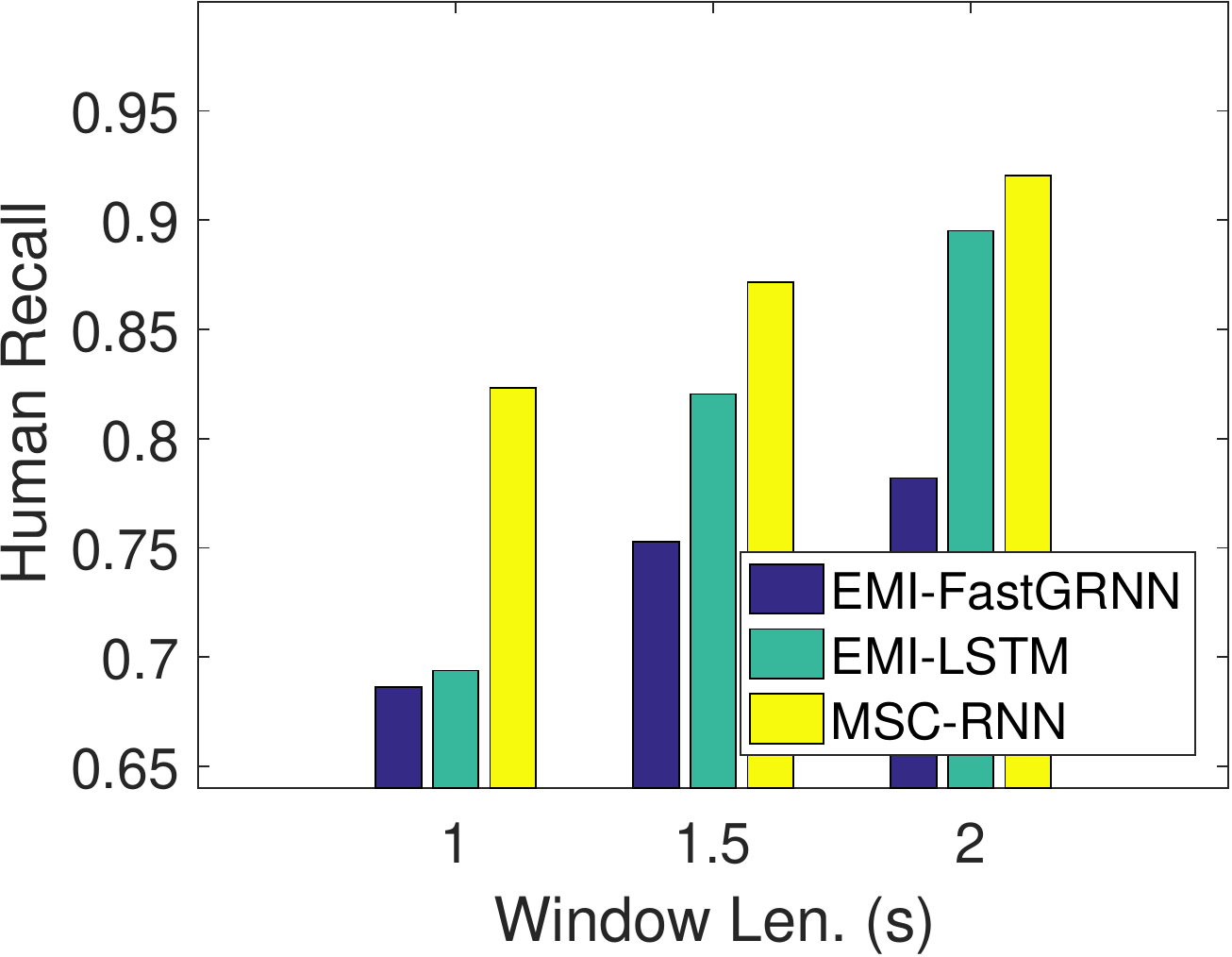}
		\caption{Human Recall (H=32)}
	\end{subfigure}%
	~
	\begin{subfigure}[h]{0.25\textwidth}
		\centering
		\includegraphics[width=\textwidth]{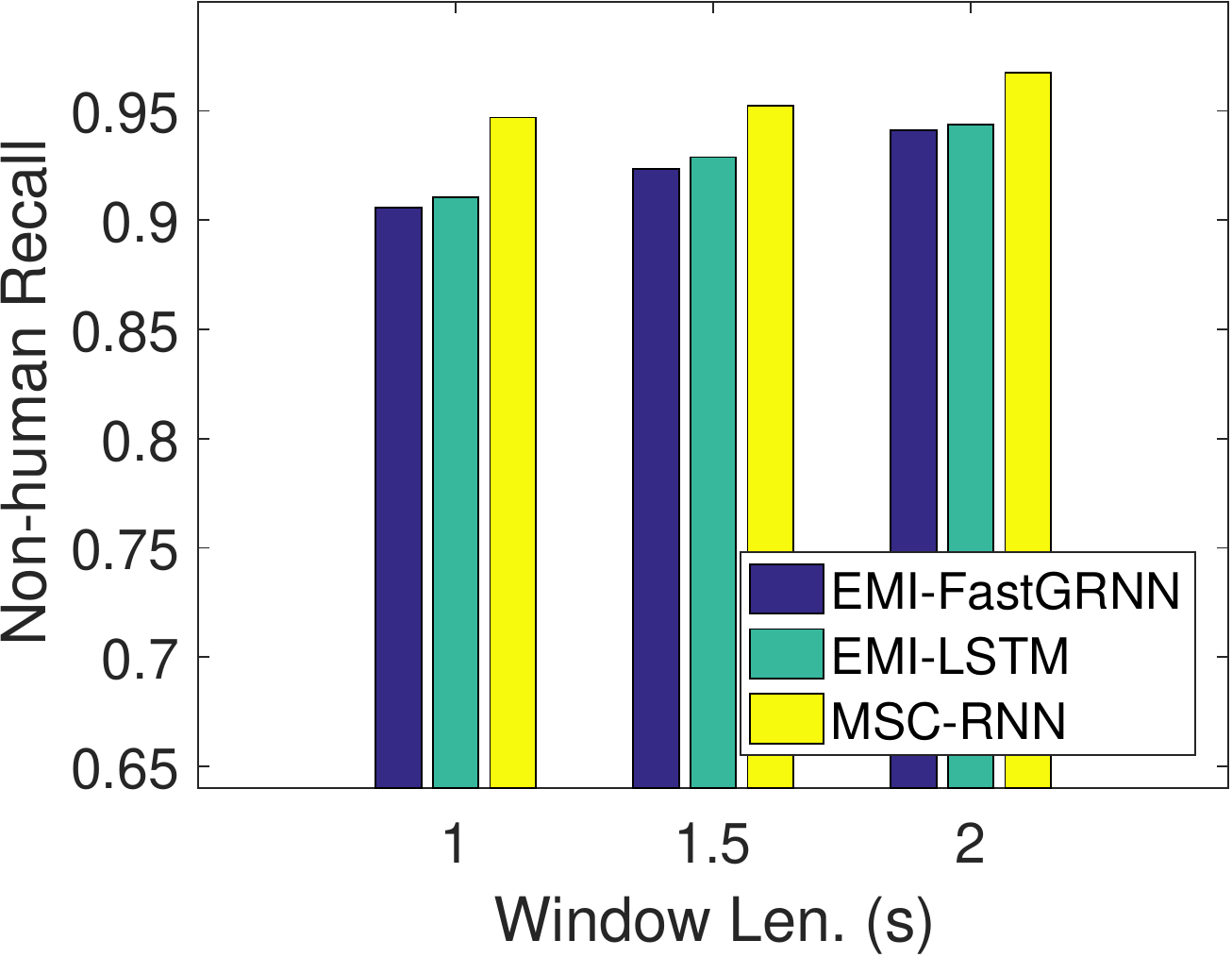}
		\caption{Non-human Recall (H=32)}
	\end{subfigure}%
	
	\vspace{5mm}
	\begin{subfigure}[h]{0.25\textwidth}
		\centering
		\includegraphics[width=\textwidth]{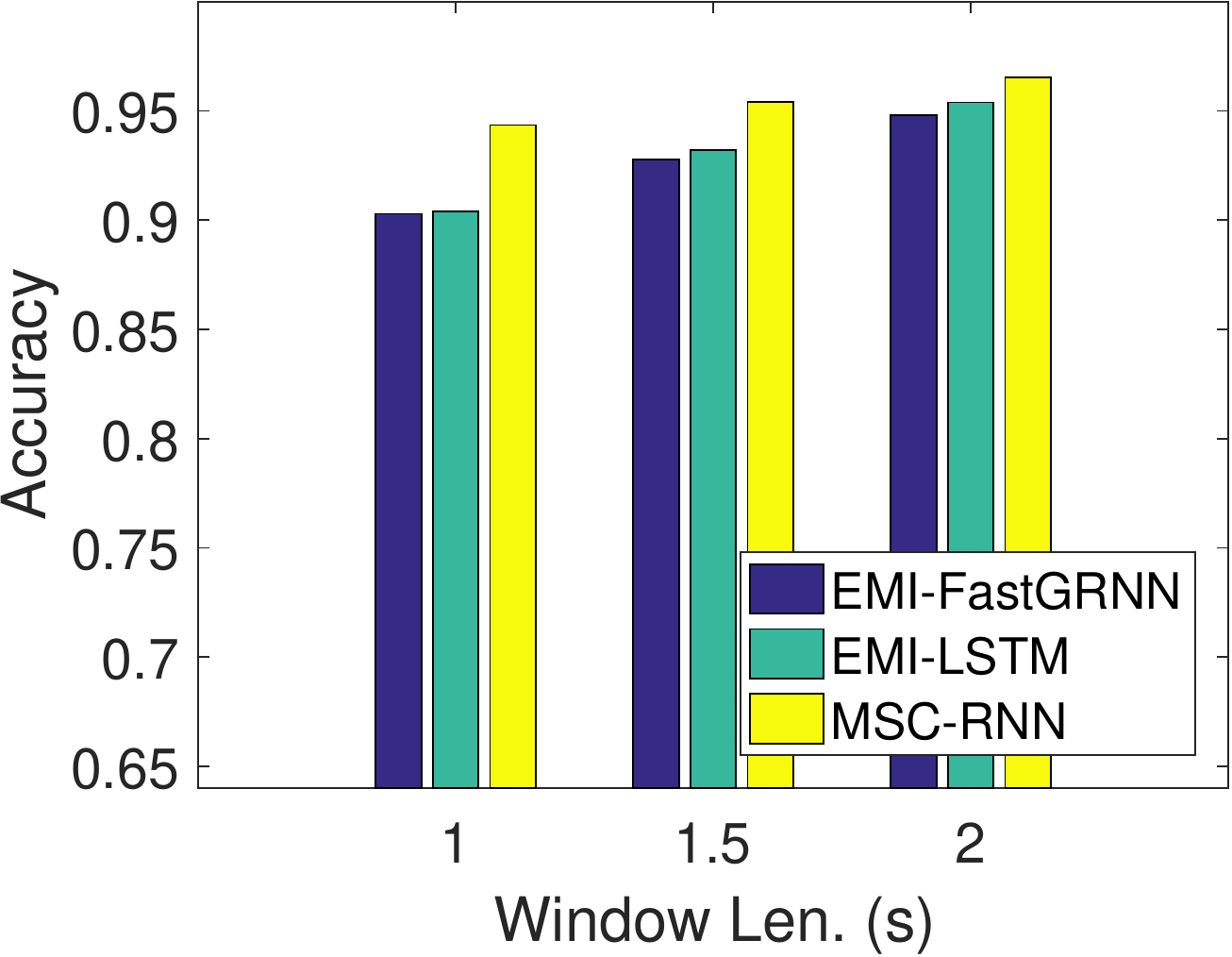}
		\caption{Accuracy (H=64)}
	\end{subfigure}%
	~
	\begin{subfigure}[h]{0.25\textwidth}
		\centering
		\includegraphics[width=\textwidth]{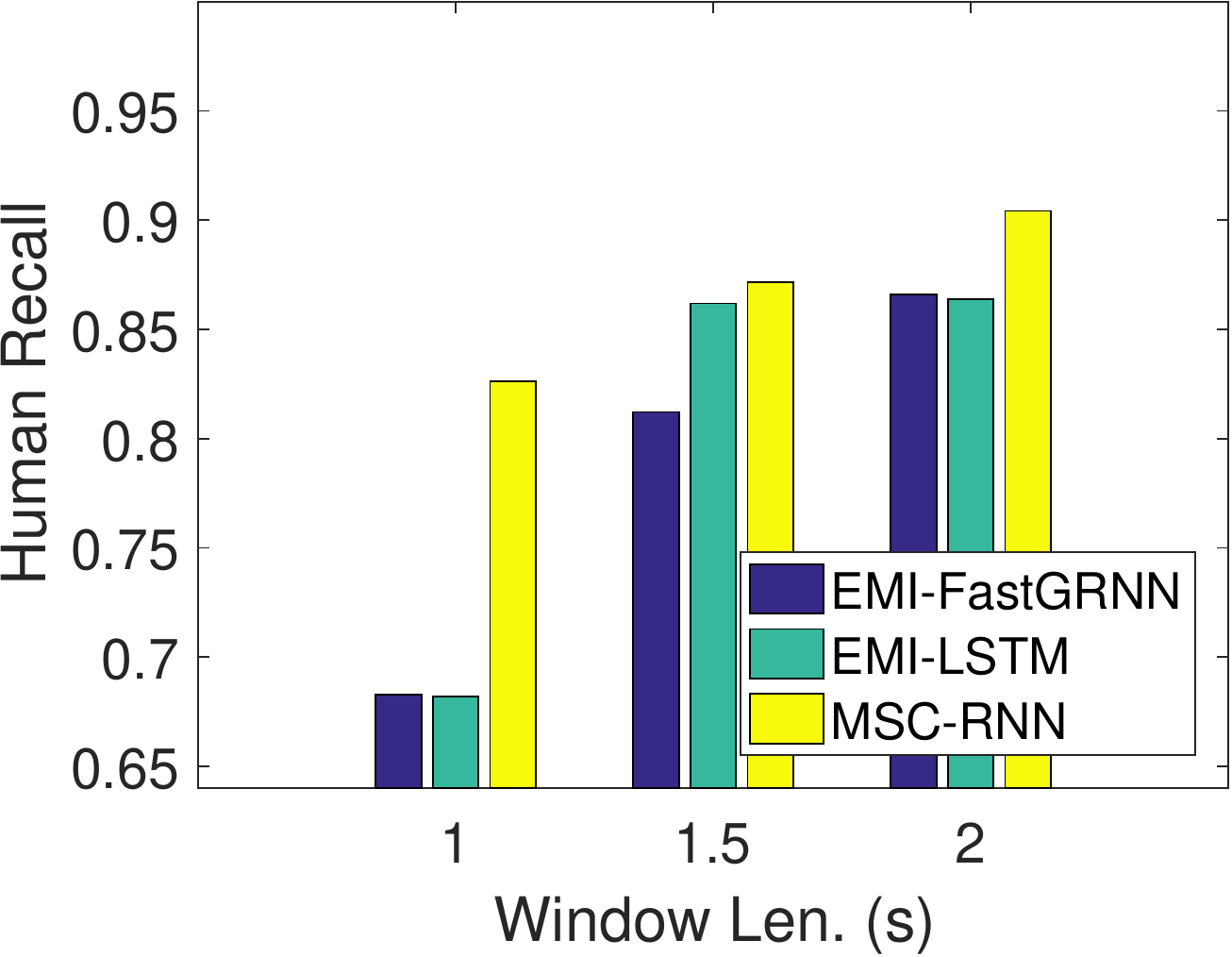}
		\caption{Human Recall (H=64)}
	\end{subfigure}%
	~
	\begin{subfigure}[h]{0.25\textwidth}
		\centering
		\includegraphics[width=\textwidth]{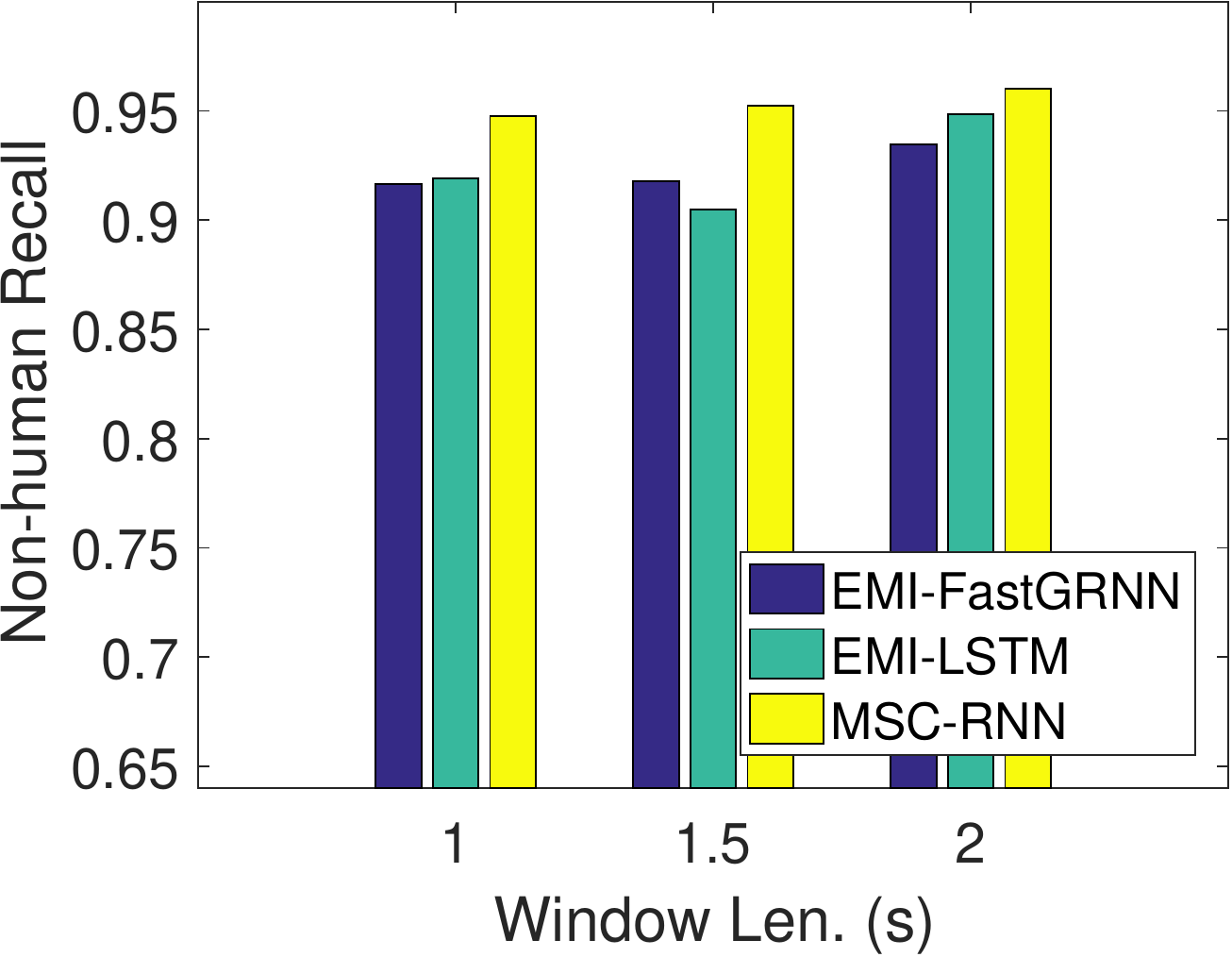}
		\caption{Non-human Recall (H=64)}
	\end{subfigure}%
	\caption{Sensing performance comparison of MSC-RNN with EMI-FastGRNN and EMI-LSTM}
	\label{fig_emi_results}
\end{figure*}

Figure \ref{fig_emi_results} contrasts our model with 3-class EMI-FastGRNN and EMI-LSTM, for fixed hidden sizes of 16, 32, and 64 respectively. It can be seen that MSC-RNN outperforms the monolithic EMI algorithms on all three metrics of accuracy, non-human and human recalls (with one exception for EMI-LSTM). Notably, cascading significantly enhances the non-dominant class recall over the other methods, especially for larger hidden sizes, and therefore offers better resilience to the source type imbalance in radar datasets.

\subsubsection{Runtime Efficiency Comparison - MSC-RNN vs. Feature Handcrafting.}
Table \ref{tab_duty_cycle} lists the runtime duty cycle estimates of MSC-RNN versus shallow SVM alternatives in two deployment contexts with realistic clutter conditions, supported by usage statistics of a popular biking trail in Columbus, OH \cite{trailusage}. While the 2-tier SVM understandably has the lowest duty cycle due to a cheap lower tier, it is not a competitive solution as established in Section \ref{subsec_svm_vs_msc}. The 3-class SVM, on the other hand, is dominated by the feature computation overhead. While the 48$\times$2 MSC-RNN formulation is about 1.75$\times$ as efficient as using handcrafted features, it is possible to reduce instance-level computations even further by using longer input vectors and reducing the number of iterations. As an example, MSC-RNN with a 16-dimensional input vector is 3$\times$ more efficient than feature engineering.

\begin{table}[htb]
	\caption{Estimated featurization duty cycle comparison on ARM Cortex-M3}
	\begin{tabular}{l|rr}
		\hline
		\multicolumn{1}{c|}{\multirow{2}{*}{\textbf{Architecture}}} & \multicolumn{2}{c}{\textbf{Est. Duty Cycle (Cortex-M3)}}                                   \\ \cline{2-3} 
		\multicolumn{1}{c|}{}                                       & \multicolumn{1}{c}{\textbf{97\% Clutter}} & \multicolumn{1}{c}{\textbf{98\% Clutter}} \\ \hline\hline
		MSC-RNN (Inp. dim.=2)                                                & 21.00\%                                     & 20.00\%                                     \\
		MSC-RNN (Inp. dim.=16)                                                & 10.87\%                                     & 10.7\%                                     \\ \hline
		2-Tier SVM                                                  & 2.05\%                                     & 1.7\%                                     \\
		3-Class SVM                                                 & 35.00\%                                    & 35.00\%                                    \\ \hline
	\end{tabular}
	\label{tab_duty_cycle}
\end{table}



\subsubsection{Tier-wise Evaluation}
\label{subsec_eval_components}
We next compare the lower-tier and upper-tier classifiers individually to their shallow counterparts in the 2-tier SVM solution.

\begin{figure}[t]
	\centering
	\includegraphics[width=0.25\textwidth]{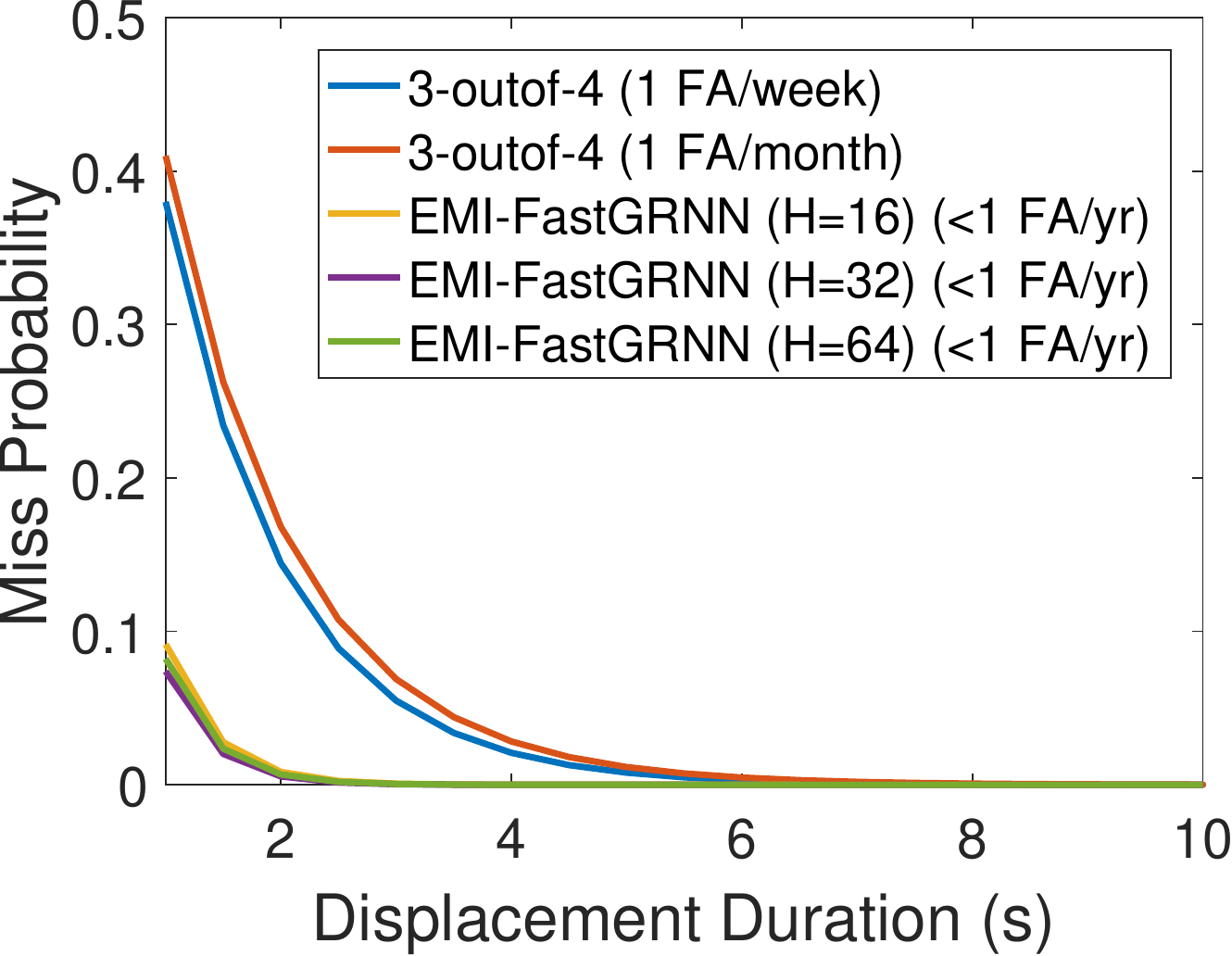}
	\caption{Comparison of miss probabilities versus displacement durations of Tier 1 classifier vs.~ 3-outof-4 phase unwrapped displacement detector (window length: 2 seconds)}
	\label{fig_missed_prob}
\end{figure}

\noindent
\textbf{Tier 1 Classifier.}\enspace Figure \ref{fig_missed_prob} compares the probabilities of missed detects versus displacement durations for the 3-outof-4 displacement detector and the EMI component of our solution with 2-second windows (for a principled approach to choosing parameters for the former, refer to Appendix \ref{sec_appendix}) at hidden sizes of 16, 32, and 64. It can be seen that, for the shortest cut length of 1.5 s in the dataset, the detection probability is improved by up to 1.5$\times$ (1.6$\times$) over the 3-outof-4 detector with false alarm rates of 1/week and 1/month respectively even when the false alarm rate ($1-$test clutter recall) of EMI is 0, which translates to a false alarm rate of $<$1 per year. Further, the EMI detector converges to 0 false detects with displacements $\geq$2.5 s, and is therefore able to reliably detect walks 2.6$\times$ shorter than the previous solution. Therefore, it is possible to restrict false positives much below 1/month while significantly improving detectability over the M-outof-N solution. Since the clutter and source datasets span various backgrounds (Figure \ref{fig_locations}), MSC-RNN offers superior cross-environmental robustness.

\vspace{1mm}
\noindent
\textbf{Tier 2 Classifier.}\enspace We now show that the gains of MSC-RNN over the 2-tier SVM solution are not, in fact, contingent on the quality of the underlying displacement detector for the latter. For this experiment, we train a 2-class FastGRNN on embeddings derived from the lower-layer EMI-FastGRNN. Table \ref{tab_twoclass} compares its performance with the upper-tier SVM from the latter when trained with the best 15 cross-domain features obtained from the raw radar samples. It can be seen that the purely time-domain FastGRNN still generally outperforms the 2-class SVM on all three metrics of accuracy, human recall, and non-human recall. Thus, it is possible to replace feature engineering with deep feature learning and enjoy the dual benefits of improved sensing and runtime efficiency for this class of radar applications.

\begin{table}[htb]
	\caption{Independent of the Tier 1 classifier, the Tier 2 source-type classifier outperforms the SVM}
	\begin{tabular}{l|cc|cc|cc}
		\hline
		\multicolumn{1}{c|}{\multirow{2}{*}{\textbf{\begin{tabular}[c]{@{}c@{}}Win.\\ Len.\\ (s)\end{tabular}}}} & \multicolumn{2}{c|}{\textbf{Accuracy}}                                     & \multicolumn{2}{c|}{\textbf{\begin{tabular}[c]{@{}c@{}}Human\\ Recall\end{tabular}}}  & \multicolumn{2}{c}{\textbf{\begin{tabular}[c]{@{}c@{}}Non-human\\ Recall\end{tabular}}}     \\ \cline{2-7} 
		\multicolumn{1}{c|}{}                                                                                    & \multicolumn{1}{c}{\textbf{\begin{tabular}[c]{@{}c@{}}SVM\\ \_15f\end{tabular}}} & \multicolumn{1}{c|}{\textbf{\begin{tabular}[c]{@{}c@{}}Fast-\\ GRNN\end{tabular}}} & \multicolumn{1}{c}{\textbf{\begin{tabular}[c]{@{}c@{}}SVM\\ \_15f\end{tabular}}} & \multicolumn{1}{c|}{\textbf{\begin{tabular}[c]{@{}c@{}}Fast-\\ GRNN\end{tabular}}} & \multicolumn{1}{c}{\textbf{\begin{tabular}[c]{@{}c@{}}SVM\\ \_15f\end{tabular}}} & \multicolumn{1}{c}{\textbf{\begin{tabular}[c]{@{}c@{}}Fast-\\ GRNN\end{tabular}}} \\ \hline\hline
		1     & 0.93    & 0.93   & 0.90   & 0.90    & 0.93   & 0.94                                                                            \\
		1.5   & 0.93    & 0.93   & 0.90   & 0.93    & 0.95   & 0.95                                                                            \\
		2     & 0.93    & 0.96   & 0.86   & 0.96    & 0.96   & 0.97                                                                            \\\hline
	\end{tabular}
	\label{tab_twoclass}
\end{table}

%% file: discussion.tex
\section{Low-power Implementation}
\label{sec_disc}

The radar sensor described in Figure \ref{fig_radar}(a) uses an ARM Cortex-M3 microcontroller with 96 KB of RAM and 4 MB of flash storage. It runs eMote \cite{emote}, a low-jitter near real-time operating system with a small footprint. We emphasize that energy efficient compute, not working memory or storage, is the bigger concern for efficient real-time operation. Hence, we take several measures to efficiently implement the multi-scale RNN to run at a low duty cycle on the device. These include low-rank representation of hidden states, Q15 quantization, and \textit{piecewise-linear} approximations of non-linear functions. The latter in particular ensures that all the computations can be performed with integer arithmetic when the weights and inputs are quantized. For example, $\tanh(x)$ can be approximated as: $quantTanh(x) = \max(\min(x, 1), -1)$, and $sigmoid(x)$ can be approximated as: $quantSigm(x) = \max(\min(\frac{x+1}{2}, 1), 0)$. The underlying linear algebraic operations are implemented using the CMSIS-DSP library \cite{cmsisdsp}. While advanced ARM processors such as Cortex-M4 offer floating point support, it should be noted that, for efficiency reasons, using sparse, low rank matrices and quantization techniques are beneficial in general.

%% file: conclusion.tex
\section{Conclusion and Future Work}
\label{sec_conc}
In this work, we introduce multi-scale, cascaded RNNs for radar sensing, and show how leveraging the ontological decomposition of a canonical classification problem into clutter vs.~source classification, followed by source type discrimination on an on-demand basis can improve both sensing quality as well as runtime efficiency over alternative systems. Learning discriminators at the time-scales relevant to their respective tasks, and jointly training the discriminators while being cognizant of the cascading behavior between them yields the desired improvement.

The extension of MSC-RNNs to more complicated sensing contexts is a topic of future work. Of interest are regression-based radar ``counting'' problems such as occupancy estimation or active transportation monitoring, where the competitiveness of MSC-RNN to architectures such as TCNs \cite{bai2018empirical} could be insightful. We also believe that MSC-RNN could also apply to alternative sensing for smart cities and built environments where the sources have intrinsic ontological hierarchies, such as in urban sound classification \cite{Bello2019}.

%% file: acknowledgement.tex
\section*{Acknowledgements}
We thank our shepherd, Zheng Yang, and the anonymous reviewers for their comments. We are indebted to Don Dennis, Prateek Jain, and Harsha Vardhan Simhadri at Microsoft Research India for their suggestions and feedback. The computation for this work was supported by the Ohio Supercomputer Center \cite{OhioSupercomputerCenter1987} project PAS1090, the IIT Delhi HPC facility, and Azure services provided by Microsoft Research Summer Workshop 2018: Machine Learning on Constrained Devices \footnote{https://www.microsoft.com/en-us/research/event/microsoft-research-summer-workshop-2018-machine-learning-on-constrained-devices/}.

%% file: appendix.tex
\appendix
\section{Parameter Selection for M-outof-N Displacement Detector}
\begin{figure}[htb]
	\begin{subfigure}[t]{0.235\textwidth}
		\includegraphics[width=\textwidth]{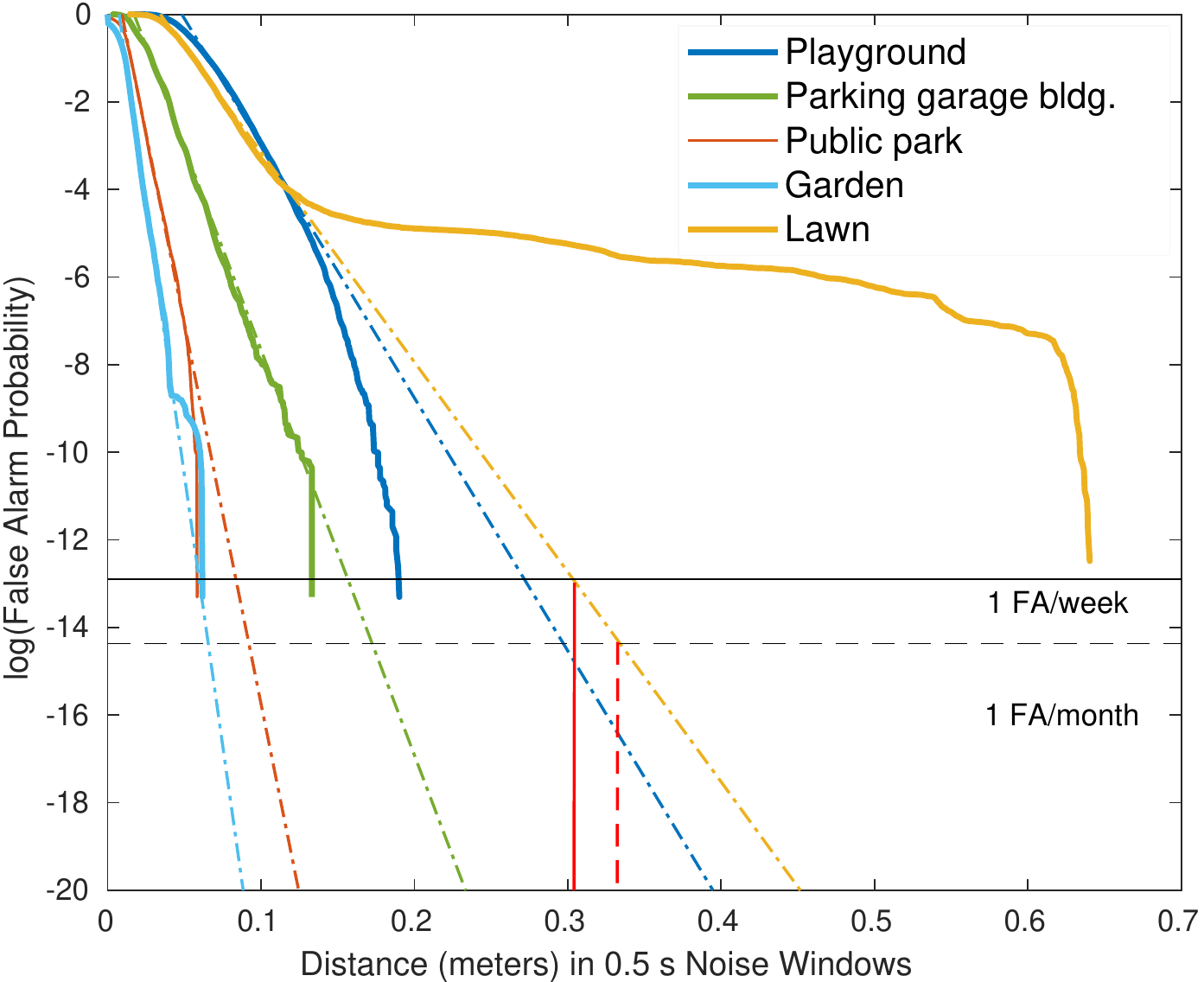}
		\caption{Clutter threshold selection for 1 FA/week and 1 FA/month}
	\end{subfigure}
	~
	\begin{subfigure}[t]{0.235\textwidth}
		\includegraphics[width=\textwidth]{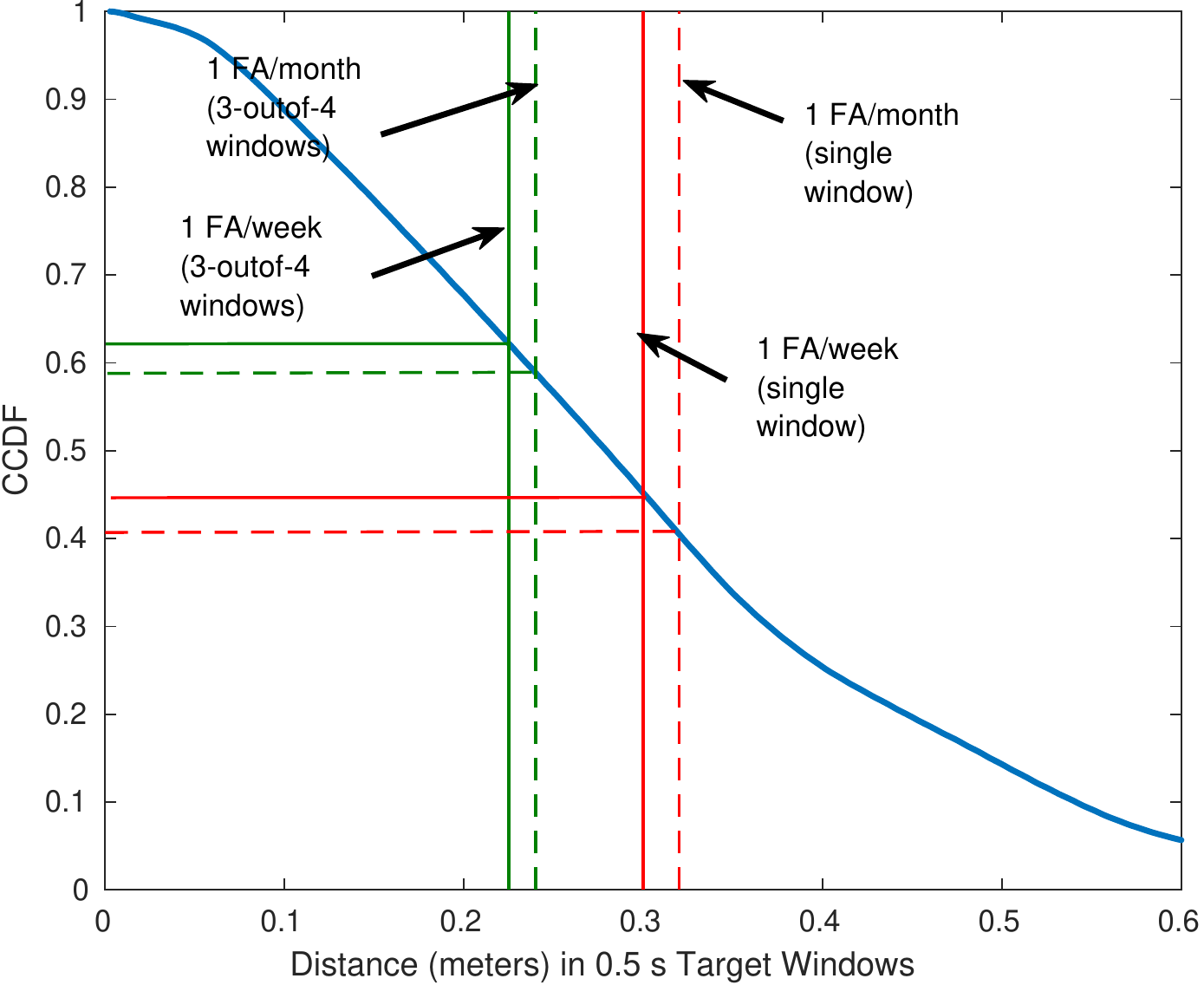}
		\caption{Relaxation of per-window threshold through aggregation}
	\end{subfigure}
	\caption{Shallow displacement detector parameter selection using the datasets from Table \ref{tab_datasets_cuts}: here, M=3 and N=4}
	\label{fig_moutofn_params}
\end{figure}
\label{sec_appendix}
We discuss the parameter selection process for the unwrapped-phase displacement detector \cite{roy2017cross} referenced in Figures \ref{fig_shallow_vs_deep} and \ref{fig_missed_prob} in a principled manner. Figure \ref{fig_moutofn_params}(a) shows the cumulative distribution of unwrapped phase changes of environmental clutter, translated into real distance units, in various environments for $\frac{1}{2}$ second integration windows from the clutter datasets in Table \ref{tab_datasets_cuts}. The data is extrapolated using linear fitting on a logarithmic scale to estimate the required phase thresholds to satisfy false alarm rates of $1$ per week and $1$ per month respectively (derived using Bernoulli probabilities). We see that the unwrapped thresholds for 1 false alarm per week and month correspond to 0.3 and 0.32 m respectively. In this analysis, we fix the IQ rejection parameter at 0.9, which gives us the most lenient thresholds.

Figure \ref{fig_moutofn_params}(b) illustrates the CCDFs of phase displacements for all target types (humans, gym balls, dogs, cattle, and slow-moving vehicles) in our dataset combined, calculated over $\frac{1}{2}$ second windows. Setting thresholds based on the previous analysis, the probability of false negatives per window is still significant. In practice, the algorithm improves detection by basing its decision on $3$-outof-$4$ sliding windows, where the miss probability improves since the threshold per window is now $\frac{3}{4}\times$ the original threshold. For $1$ false alarm per week (month), the displacement threshold for the $3$-outof-$4$ detector reduces to 0.22 m (0.24 m) per window, with an improved miss probability of 0.59 (0.62).

%% file: main.bbl

\begin{thebibliography}{47}


\ifx \showCODEN    \undefined \def \showCODEN     #1{\unskip}     \fi
\ifx \showDOI      \undefined \def \showDOI       #1{#1}\fi
\ifx \showISBNx    \undefined \def \showISBNx     #1{\unskip}     \fi
\ifx \showISBNxiii \undefined \def \showISBNxiii  #1{\unskip}     \fi
\ifx \showISSN     \undefined \def \showISSN      #1{\unskip}     \fi
\ifx \showLCCN     \undefined \def \showLCCN      #1{\unskip}     \fi
\ifx \shownote     \undefined \def \shownote      #1{#1}          \fi
\ifx \showarticletitle \undefined \def \showarticletitle #1{#1}   \fi
\ifx \showURL      \undefined \def \showURL       {\relax}        \fi
\providecommand\bibfield[2]{#2}
\providecommand\bibinfo[2]{#2}
\providecommand\natexlab[1]{#1}
\providecommand\showeprint[2][]{arXiv:#2}

\bibitem[\protect\citeauthoryear{??}{cms}{[n. d.]}]%
        {cmsisdsp}
 \bibinfo{year}{[n. d.]}\natexlab{}.
\newblock \bibinfo{title}{CMSIS-DSP Software Library}.
\newblock
  \bibinfo{howpublished}{\url{http://www.keil.com/pack/doc/CMSIS/DSP/html/index.html}}.
\newblock


\bibitem[\protect\citeauthoryear{??}{tra}{[n. d.]}]%
        {trailusage}
 \bibinfo{year}{[n. d.]}\natexlab{}.
\newblock \bibinfo{title}{Olentangy Trail usage, Columbus, OH}.
\newblock
  \bibinfo{howpublished}{\url{https://www.columbus.gov/recreationandparks/trails/Future-Trails-(Updated)/}}.
\newblock


\bibitem[\protect\citeauthoryear{Adkins, Ghena, et~al\mbox{.}}{Adkins
  et~al\mbox{.}}{2018}]%
        {adkins2018signpost}
\bibfield{author}{\bibinfo{person}{Joshua Adkins}, \bibinfo{person}{Branden
  Ghena}, {et~al\mbox{.}}} \bibinfo{year}{2018}\natexlab{}.
\newblock \showarticletitle{The signpost platform for city-scale sensing}. In
  \bibinfo{booktitle}{\emph{IPSN}}. \bibinfo{publisher}{IEEE},
  \bibinfo{pages}{188--199}.
\newblock


\bibitem[\protect\citeauthoryear{Bai, Kolter, et~al\mbox{.}}{Bai
  et~al\mbox{.}}{2018}]%
        {bai2018empirical}
\bibfield{author}{\bibinfo{person}{Shaojie Bai}, \bibinfo{person}{J~Zico
  Kolter}, {et~al\mbox{.}}} \bibinfo{year}{2018}\natexlab{}.
\newblock \showarticletitle{An empirical evaluation of generic convolutional
  and recurrent networks for sequence modeling}.
\newblock \bibinfo{journal}{\emph{arXiv preprint arXiv:1803.01271}}
  (\bibinfo{year}{2018}).
\newblock


\bibitem[\protect\citeauthoryear{Bello, Silva, et~al\mbox{.}}{Bello
  et~al\mbox{.}}{2019}]%
        {Bello2019}
\bibfield{author}{\bibinfo{person}{Juan~P. Bello}, \bibinfo{person}{Claudio
  Silva}, {et~al\mbox{.}}} \bibinfo{year}{2019}\natexlab{}.
\newblock \showarticletitle{SONYC: A system for monitoring, analyzing, and
  mitigating urban noise pollution}.
\newblock \bibinfo{journal}{\emph{CACM}} \bibinfo{volume}{62},
  \bibinfo{number}{2} (\bibinfo{date}{Jan.} \bibinfo{year}{2019}),
  \bibinfo{pages}{68--77}.
\newblock
\showISSN{0001-0782}


\bibitem[\protect\citeauthoryear{Cand{\`e}s, Li, et~al\mbox{.}}{Cand{\`e}s
  et~al\mbox{.}}{2011}]%
        {candes2011robust}
\bibfield{author}{\bibinfo{person}{Emmanuel~J Cand{\`e}s},
  \bibinfo{person}{Xiaodong Li}, {et~al\mbox{.}}}
  \bibinfo{year}{2011}\natexlab{}.
\newblock \showarticletitle{Robust principal component analysis?}
\newblock \bibinfo{journal}{\emph{JACM}} \bibinfo{volume}{58},
  \bibinfo{number}{3} (\bibinfo{year}{2011}), \bibinfo{pages}{11}.
\newblock


\bibitem[\protect\citeauthoryear{Catlett, Beckman, et~al\mbox{.}}{Catlett
  et~al\mbox{.}}{2017}]%
        {catlett2017array}
\bibfield{author}{\bibinfo{person}{Charles~E Catlett}, \bibinfo{person}{Peter~H
  Beckman}, {et~al\mbox{.}}} \bibinfo{year}{2017}\natexlab{}.
\newblock \showarticletitle{Array of things: A scientific research instrument
  in the public way: platform design and early lessons learned}. In
  \bibinfo{booktitle}{\emph{SCOPE}}. ACM, \bibinfo{pages}{26--33}.
\newblock


\bibitem[\protect\citeauthoryear{Center}{Center}{1987}]%
        {OhioSupercomputerCenter1987}
\bibfield{author}{\bibinfo{person}{Ohio~Supercomputer Center}.}
  \bibinfo{year}{1987}\natexlab{}.
\newblock \bibinfo{title}{Ohio Supercomputer Center}.
\newblock
\newblock
\urldef\tempurl%
\url{http://osc.edu/ark:/19495/f5s1ph73}
\showURL{%
\tempurl}


\bibitem[\protect\citeauthoryear{Cho, Van~Merri{\"e}nboer, et~al\mbox{.}}{Cho
  et~al\mbox{.}}{2014}]%
        {cho2014properties}
\bibfield{author}{\bibinfo{person}{Kyunghyun Cho}, \bibinfo{person}{Bart
  Van~Merri{\"e}nboer}, {et~al\mbox{.}}} \bibinfo{year}{2014}\natexlab{}.
\newblock \showarticletitle{On the properties of neural machine translation:
  Encoder-decoder approaches}.
\newblock \bibinfo{journal}{\emph{arXiv preprint arXiv:1409.1259}}
  (\bibinfo{year}{2014}).
\newblock


\bibitem[\protect\citeauthoryear{Chung, Ahn, et~al\mbox{.}}{Chung
  et~al\mbox{.}}{2016}]%
        {chung2016hierarchical}
\bibfield{author}{\bibinfo{person}{Junyoung Chung}, \bibinfo{person}{Sungjin
  Ahn}, {et~al\mbox{.}}} \bibinfo{year}{2016}\natexlab{}.
\newblock \showarticletitle{Hierarchical multiscale recurrent neural networks}.
\newblock \bibinfo{journal}{\emph{arXiv preprint arXiv:1609.01704}}
  (\bibinfo{year}{2016}).
\newblock


\bibitem[\protect\citeauthoryear{Dennis, Pabbaraju, et~al\mbox{.}}{Dennis
  et~al\mbox{.}}{2018}]%
        {dennis2018multiple}
\bibfield{author}{\bibinfo{person}{Don Dennis}, \bibinfo{person}{Chirag
  Pabbaraju}, {et~al\mbox{.}}} \bibinfo{year}{2018}\natexlab{}.
\newblock \showarticletitle{Multiple instance learning for efficient sequential
  data classification on resource-constrained devices}. In
  \bibinfo{booktitle}{\emph{NIPS}}. \bibinfo{publisher}{Curran Associates,
  Inc}, \bibinfo{pages}{10975--10986}.
\newblock


\bibitem[\protect\citeauthoryear{Dennis, Gopinath, et~al\mbox{.}}{Dennis
  et~al\mbox{.}}{[n. d.]}]%
        {edgemlcode}
\bibfield{author}{\bibinfo{person}{Don~Kurian Dennis}, \bibinfo{person}{Sridhar
  Gopinath}, {et~al\mbox{.}}} \bibinfo{year}{[n. d.]}\natexlab{}.
\newblock \bibinfo{title}{{EdgeML: Machine learning for resource-constrained
  edge devices}}.
\newblock
\newblock
\urldef\tempurl%
\url{https://github.com/Microsoft/EdgeML}
\showURL{%
\tempurl}


\bibitem[\protect\citeauthoryear{Goldstein, Zebker, et~al\mbox{.}}{Goldstein
  et~al\mbox{.}}{1988}]%
        {goldstein1988satellite}
\bibfield{author}{\bibinfo{person}{Richard~M Goldstein},
  \bibinfo{person}{Howard~A Zebker}, {et~al\mbox{.}}}
  \bibinfo{year}{1988}\natexlab{}.
\newblock \showarticletitle{Satellite radar interferometry: Two-dimensional
  phase unwrapping}.
\newblock \bibinfo{journal}{\emph{Radio science}} \bibinfo{volume}{23},
  \bibinfo{number}{4} (\bibinfo{year}{1988}), \bibinfo{pages}{713--720}.
\newblock


\bibitem[\protect\citeauthoryear{Han, Mao, et~al\mbox{.}}{Han
  et~al\mbox{.}}{2016}]%
        {han2015deep_compression}
\bibfield{author}{\bibinfo{person}{Song Han}, \bibinfo{person}{Huizi Mao},
  {et~al\mbox{.}}} \bibinfo{year}{2016}\natexlab{}.
\newblock \showarticletitle{Deep compression: Compressing deep neural networks
  with pruning, trained quantization and Huffman coding}.
\newblock \bibinfo{journal}{\emph{ICLR}} (\bibinfo{year}{2016}).
\newblock


\bibitem[\protect\citeauthoryear{He and Arora}{He and Arora}{2014}]%
        {he2014regression}
\bibfield{author}{\bibinfo{person}{Jin He} {and} \bibinfo{person}{Anish
  Arora}.} \bibinfo{year}{2014}\natexlab{}.
\newblock \showarticletitle{A regression-based radar-mote system for people
  counting}. In \bibinfo{booktitle}{\emph{PerCom}}. \bibinfo{publisher}{IEEE},
  \bibinfo{pages}{95--102}.
\newblock


\bibitem[\protect\citeauthoryear{He, Roy, et~al\mbox{.}}{He
  et~al\mbox{.}}{2014}]%
        {he2014mote}
\bibfield{author}{\bibinfo{person}{Jin He}, \bibinfo{person}{Dhrubojyoti Roy},
  {et~al\mbox{.}}} \bibinfo{year}{2014}\natexlab{}.
\newblock \showarticletitle{Mote-scale human-animal classification via
  micropower radar}. In \bibinfo{booktitle}{\emph{SenSys}}.
  \bibinfo{publisher}{ACM}, \bibinfo{pages}{328--329}.
\newblock


\bibitem[\protect\citeauthoryear{Hinton}{Hinton}{1990}]%
        {hinton1990mapping}
\bibfield{author}{\bibinfo{person}{Geoffrey~E Hinton}.}
  \bibinfo{year}{1990}\natexlab{}.
\newblock \showarticletitle{Mapping part-whole hierarchies into connectionist
  networks}.
\newblock \bibinfo{journal}{\emph{Artificial Intelligence}}
  \bibinfo{volume}{46}, \bibinfo{number}{1-2} (\bibinfo{year}{1990}),
  \bibinfo{pages}{47--75}.
\newblock


\bibitem[\protect\citeauthoryear{Hochreiter and Schmidhuber}{Hochreiter and
  Schmidhuber}{1997}]%
        {hochreiter1997long}
\bibfield{author}{\bibinfo{person}{Sepp Hochreiter} {and}
  \bibinfo{person}{J{\"u}rgen Schmidhuber}.} \bibinfo{year}{1997}\natexlab{}.
\newblock \showarticletitle{Long short-term memory}.
\newblock \bibinfo{journal}{\emph{Neural computation}} \bibinfo{volume}{9},
  \bibinfo{number}{8} (\bibinfo{year}{1997}), \bibinfo{pages}{1735--1780}.
\newblock


\bibitem[\protect\citeauthoryear{James, Witten, et~al\mbox{.}}{James
  et~al\mbox{.}}{2013}]%
        {james2013introduction}
\bibfield{author}{\bibinfo{person}{Gareth James}, \bibinfo{person}{Daniela
  Witten}, {et~al\mbox{.}}} \bibinfo{year}{2013}\natexlab{}.
\newblock \bibinfo{booktitle}{\emph{An introduction to statistical learning}}.
  Vol.~\bibinfo{volume}{112}.
\newblock \bibinfo{publisher}{Springer}.
\newblock


\bibitem[\protect\citeauthoryear{Jesus~Javier and Kim}{Jesus~Javier and
  Kim}{2014}]%
        {lpc}
\bibfield{author}{\bibinfo{person}{Rios Jesus~Javier} {and}
  \bibinfo{person}{Youngwook Kim}.} \bibinfo{year}{2014}\natexlab{}.
\newblock \showarticletitle{Application of linear predictive coding for human
  activity classification based on micro-Doppler signatures}.
\newblock \bibinfo{journal}{\emph{GRSL, IEEE}}  \bibinfo{volume}{11}
  (\bibinfo{date}{10} \bibinfo{year}{2014}), \bibinfo{pages}{1831--1834}.
\newblock


\bibitem[\protect\citeauthoryear{Jokanovic, Amin, et~al\mbox{.}}{Jokanovic
  et~al\mbox{.}}{2016}]%
        {fallDetectionDNN}
\bibfield{author}{\bibinfo{person}{Branka Jokanovic}, \bibinfo{person}{Moeness
  Amin}, {et~al\mbox{.}}} \bibinfo{year}{2016}\natexlab{}.
\newblock \showarticletitle{Radar fall motion detection using deep learning}.
  In \bibinfo{booktitle}{\emph{RADAR}}. \bibinfo{publisher}{IEEE},
  \bibinfo{pages}{1--6}.
\newblock
\showISSN{2375-5318}


\bibitem[\protect\citeauthoryear{Jose, Cisse, et~al\mbox{.}}{Jose
  et~al\mbox{.}}{2017}]%
        {jose2017kronecker}
\bibfield{author}{\bibinfo{person}{Cijo Jose}, \bibinfo{person}{Moustpaha
  Cisse}, {et~al\mbox{.}}} \bibinfo{year}{2017}\natexlab{}.
\newblock \showarticletitle{Kronecker recurrent units}.
\newblock \bibinfo{journal}{\emph{arXiv preprint arXiv:1705.10142}}
  (\bibinfo{year}{2017}).
\newblock


\bibitem[\protect\citeauthoryear{Kim and Ling}{Kim and Ling}{2008}]%
        {kim2008human}
\bibfield{author}{\bibinfo{person}{Youngwook Kim} {and} \bibinfo{person}{Hao
  Ling}.} \bibinfo{year}{2008}\natexlab{}.
\newblock \showarticletitle{Human activity classification based on
  micro-Doppler signatures using an artificial neural network}. In
  \bibinfo{booktitle}{\emph{AP-S}}. \bibinfo{publisher}{IEEE},
  \bibinfo{pages}{1--4}.
\newblock


\bibitem[\protect\citeauthoryear{Kim and Moon}{Kim and Moon}{2015}]%
        {kim2015humandetCNN}
\bibfield{author}{\bibinfo{person}{Youngwook Kim} {and} \bibinfo{person}{Taesup
  Moon}.} \bibinfo{year}{2015}\natexlab{}.
\newblock \showarticletitle{Human detection and activity classification based
  on micro-Doppler signatures using deep convolutional neural networks}.
\newblock \bibinfo{journal}{\emph{GRSL, IEEE}}  \bibinfo{volume}{13}
  (\bibinfo{date}{11} \bibinfo{year}{2015}), \bibinfo{pages}{1--5}.
\newblock


\bibitem[\protect\citeauthoryear{Kim and Toomajian}{Kim and Toomajian}{2016}]%
        {kim2016gestureCNN}
\bibfield{author}{\bibinfo{person}{Youngwook Kim} {and} \bibinfo{person}{Brian
  Toomajian}.} \bibinfo{year}{2016}\natexlab{}.
\newblock \showarticletitle{Hand gesture recognition using micro-Doppler
  signatures with convolutional neural network}.
\newblock \bibinfo{journal}{\emph{IEEE Access}}  \bibinfo{volume}{4}
  (\bibinfo{date}{01} \bibinfo{year}{2016}), \bibinfo{pages}{1--1}.
\newblock


\bibitem[\protect\citeauthoryear{Kotzias, Denil, et~al\mbox{.}}{Kotzias
  et~al\mbox{.}}{2014}]%
        {kotzias2014deep}
\bibfield{author}{\bibinfo{person}{Dimitrios Kotzias}, \bibinfo{person}{Misha
  Denil}, {et~al\mbox{.}}} \bibinfo{year}{2014}\natexlab{}.
\newblock \showarticletitle{Deep multi-instance transfer learning}.
\newblock \bibinfo{journal}{\emph{arXiv preprint arXiv:1411.3128}}
  (\bibinfo{year}{2014}).
\newblock


\bibitem[\protect\citeauthoryear{Krizhevsky, Sutskever,
  et~al\mbox{.}}{Krizhevsky et~al\mbox{.}}{2012}]%
        {krizhevsky2012imagenet}
\bibfield{author}{\bibinfo{person}{Alex Krizhevsky}, \bibinfo{person}{Ilya
  Sutskever}, {et~al\mbox{.}}} \bibinfo{year}{2012}\natexlab{}.
\newblock \showarticletitle{Imagenet classification with deep convolutional
  neural networks}. In \bibinfo{booktitle}{\emph{NIPS}}.
  \bibinfo{pages}{1097--1105}.
\newblock


\bibitem[\protect\citeauthoryear{Kumar, Goyal, et~al\mbox{.}}{Kumar
  et~al\mbox{.}}{2017}]%
        {kumar2017resource}
\bibfield{author}{\bibinfo{person}{Ashish Kumar}, \bibinfo{person}{Saurabh
  Goyal}, {et~al\mbox{.}}} \bibinfo{year}{2017}\natexlab{}.
\newblock \showarticletitle{Resource-efficient machine learning in 2 KB RAM for
  the Internet of Things}. In \bibinfo{booktitle}{\emph{ICML}}.
  \bibinfo{publisher}{ACM}, \bibinfo{pages}{1935--1944}.
\newblock


\bibitem[\protect\citeauthoryear{Kumari, Roy, et~al\mbox{.}}{Kumari
  et~al\mbox{.}}{2019}]%
        {kumari2019edgel}
\bibfield{author}{\bibinfo{person}{Sangeeta Kumari},
  \bibinfo{person}{Dhrubojyoti Roy}, {et~al\mbox{.}}}
  \bibinfo{year}{2019}\natexlab{}.
\newblock \showarticletitle{EdgeL\textsuperscript{3}: Compressing
  L\textsuperscript{3}-Net for mote scale urban noise monitoring}. In
  \bibinfo{booktitle}{\emph{IPDPSW}}. IEEE, \bibinfo{pages}{877--884}.
\newblock


\bibitem[\protect\citeauthoryear{Kusupati, Singh, et~al\mbox{.}}{Kusupati
  et~al\mbox{.}}{2018}]%
        {kusupati2018fastgrnn}
\bibfield{author}{\bibinfo{person}{Aditya Kusupati}, \bibinfo{person}{Manish
  Singh}, {et~al\mbox{.}}} \bibinfo{year}{2018}\natexlab{}.
\newblock \showarticletitle{FastGRNN: A fast, accurate, stable and tiny
  kilobyte sized gated recurrent neural network}. In
  \bibinfo{booktitle}{\emph{NIPS}}. \bibinfo{publisher}{Curran Associates,
  Inc.}, \bibinfo{pages}{9030--9041}.
\newblock


\bibitem[\protect\citeauthoryear{Lam~Phung, Tivive, et~al\mbox{.}}{Lam~Phung
  et~al\mbox{.}}{2015}]%
        {logGabor}
\bibfield{author}{\bibinfo{person}{Son Lam~Phung}, \bibinfo{person}{Fok
  Hing~Chi Tivive}, {et~al\mbox{.}}} \bibinfo{year}{2015}\natexlab{}.
\newblock \showarticletitle{Classification of micro-Doppler signatures of human
  motions using log-Gabor filters}.
\newblock \bibinfo{journal}{\emph{IET Radar, Sonar \& Navigation}}
  \bibinfo{volume}{9} (\bibinfo{date}{10} \bibinfo{year}{2015}).
\newblock


\bibitem[\protect\citeauthoryear{Liu, Popescu, et~al\mbox{.}}{Liu
  et~al\mbox{.}}{2011}]%
        {fallDetection}
\bibfield{author}{\bibinfo{person}{Liang Liu}, \bibinfo{person}{Mihail
  Popescu}, {et~al\mbox{.}}} \bibinfo{year}{2011}\natexlab{}.
\newblock \showarticletitle{Automatic fall detection based on Doppler radar
  motion signature}. In \bibinfo{booktitle}{\emph{PervasiveHealth}},
  Vol.~\bibinfo{volume}{222}. \bibinfo{pages}{222--225}.
\newblock


\bibitem[\protect\citeauthoryear{Mendis, Randeny, et~al\mbox{.}}{Mendis
  et~al\mbox{.}}{2016}]%
        {uasDetection}
\bibfield{author}{\bibinfo{person}{Gihan Mendis}, \bibinfo{person}{Tharindu
  Randeny}, {et~al\mbox{.}}} \bibinfo{year}{2016}\natexlab{}.
\newblock \showarticletitle{Deep learning based doppler radar for micro UAS
  detection and classification}. In \bibinfo{booktitle}{\emph{MILCOM}}.
  \bibinfo{publisher}{IEEE}, \bibinfo{pages}{924--929}.
\newblock


\bibitem[\protect\citeauthoryear{Molchanov, Astola, et~al\mbox{.}}{Molchanov
  et~al\mbox{.}}{2011}]%
        {molchanov2011ground}
\bibfield{author}{\bibinfo{person}{Pavlo Molchanov}, \bibinfo{person}{Jaakko
  Astola}, {et~al\mbox{.}}} \bibinfo{year}{2011}\natexlab{}.
\newblock \showarticletitle{Ground moving target classification by using DCT
  coefficients extracted from micro-Doppler radar signatures and artificial
  neuron network}. In \bibinfo{booktitle}{\emph{MRRS}}.
  \bibinfo{publisher}{IEEE}, \bibinfo{pages}{173--176}.
\newblock


\bibitem[\protect\citeauthoryear{Mozer}{Mozer}{1992}]%
        {mozer1992}
\bibfield{author}{\bibinfo{person}{Michael~C Mozer}.}
  \bibinfo{year}{1992}\natexlab{}.
\newblock \showarticletitle{Induction of multiscale temporal structure}.
\newblock In \bibinfo{booktitle}{\emph{NIPS}}.
  \bibinfo{publisher}{Morgan-Kaufmann}, \bibinfo{pages}{275--282}.
\newblock


\bibitem[\protect\citeauthoryear{P.~Fairchild and Narayanan}{P.~Fairchild and
  Narayanan}{2014}]%
        {EMDecomp}
\bibfield{author}{\bibinfo{person}{Dustin P.~Fairchild} {and}
  \bibinfo{person}{Ram Narayanan}.} \bibinfo{year}{2014}\natexlab{}.
\newblock \showarticletitle{Classification of human motions using empirical
  mode decomposition of human micro-Doppler signatures}.
\newblock \bibinfo{journal}{\emph{Radar, Sonar \& Navigation, IET}}
  \bibinfo{volume}{8} (\bibinfo{date}{06} \bibinfo{year}{2014}),
  \bibinfo{pages}{425--434}.
\newblock


\bibitem[\protect\citeauthoryear{Pascanu, Mikolov, et~al\mbox{.}}{Pascanu
  et~al\mbox{.}}{2013}]%
        {pascanu2013difficulty}
\bibfield{author}{\bibinfo{person}{Razvan Pascanu}, \bibinfo{person}{Tomas
  Mikolov}, {et~al\mbox{.}}} \bibinfo{year}{2013}\natexlab{}.
\newblock \showarticletitle{On the difficulty of training recurrent neural
  networks}. In \bibinfo{booktitle}{\emph{ICML}}. \bibinfo{publisher}{ACM},
  \bibinfo{pages}{1310--1318}.
\newblock


\bibitem[\protect\citeauthoryear{Peng, Long, et~al\mbox{.}}{Peng
  et~al\mbox{.}}{2005}]%
        {peng2005feature}
\bibfield{author}{\bibinfo{person}{Hanchuan Peng}, \bibinfo{person}{Fuhui
  Long}, {et~al\mbox{.}}} \bibinfo{year}{2005}\natexlab{}.
\newblock \showarticletitle{Feature selection based on mutual information
  criteria of max-dependency, max-relevance, and min-redundancy}.
\newblock \bibinfo{journal}{\emph{IEEE Transactions on pattern analysis and
  machine intelligence}} \bibinfo{volume}{27}, \bibinfo{number}{8}
  (\bibinfo{year}{2005}), \bibinfo{pages}{1226--1238}.
\newblock


\bibitem[\protect\citeauthoryear{Roy, Kumari, et~al\mbox{.}}{Roy
  et~al\mbox{.}}{[n. d.]}]%
        {mscrnncode}
\bibfield{author}{\bibinfo{person}{Dhrubojyoti Roy}, \bibinfo{person}{Sangeeta
  Kumari}, {et~al\mbox{.}}} \bibinfo{year}{[n. d.]}\natexlab{}.
\newblock \bibinfo{title}{{MSC-RNN: Multi-Scale, Cascaded RNNs for Radar
  Classification}}.
\newblock
\newblock
\urldef\tempurl%
\url{https://github.com/dhruboroy29/MSCRNN}
\showURL{%
\tempurl}


\bibitem[\protect\citeauthoryear{Roy, Morse, et~al\mbox{.}}{Roy
  et~al\mbox{.}}{2017}]%
        {roy2017cross}
\bibfield{author}{\bibinfo{person}{Dhrubojyoti Roy},
  \bibinfo{person}{Christopher Morse}, {et~al\mbox{.}}}
  \bibinfo{year}{2017}\natexlab{}.
\newblock \showarticletitle{Cross-environmentally robust intruder
  discrimination in radar motes}. In \bibinfo{booktitle}{\emph{MASS}}.
  \bibinfo{publisher}{IEEE}, \bibinfo{pages}{426--434}.
\newblock


\bibitem[\protect\citeauthoryear{Schmidhuber}{Schmidhuber}{1992}]%
        {schmidhuber1992learning}
\bibfield{author}{\bibinfo{person}{J{\"u}rgen Schmidhuber}.}
  \bibinfo{year}{1992}\natexlab{}.
\newblock \showarticletitle{Learning complex, extended sequences using the
  principle of history compression}.
\newblock \bibinfo{journal}{\emph{Neural Computation}} \bibinfo{volume}{4},
  \bibinfo{number}{2} (\bibinfo{year}{1992}), \bibinfo{pages}{234--242}.
\newblock


\bibitem[\protect\citeauthoryear{{The Samraksh Company}}{{The Samraksh
  Company}}{[n. d.]}]%
        {emote}
\bibfield{author}{\bibinfo{person}{{The Samraksh Company}}.} \bibinfo{year}{[n.
  d.]}\natexlab{}.
\newblock \bibinfo{title}{.{NOW} with e{M}ote}.
\newblock \bibinfo{howpublished}{\url{https://goo.gl/C4CCv4}}.
\newblock


\bibitem[\protect\citeauthoryear{{UrbanCCD}}{{UrbanCCD}}{[n. d.]}]%
        {AoT}
\bibfield{author}{\bibinfo{person}{{UrbanCCD}}.} \bibinfo{year}{[n.
  d.]}\natexlab{}.
\newblock \bibinfo{title}{Array of Things}.
\newblock
  \bibinfo{howpublished}{\url{https://medium.com/array-of-things/five-years-100-nodes-and-more-to-come-d3802653db9f}}.
\newblock


\bibitem[\protect\citeauthoryear{Wang, Lin, et~al\mbox{.}}{Wang
  et~al\mbox{.}}{2017}]%
        {wang2017accelerating}
\bibfield{author}{\bibinfo{person}{Zhisheng Wang}, \bibinfo{person}{Jun Lin},
  {et~al\mbox{.}}} \bibinfo{year}{2017}\natexlab{}.
\newblock \showarticletitle{Accelerating recurrent neural networks: A
  memory-efficient approach}.
\newblock \bibinfo{journal}{\emph{IEEE Transactions on VLSI Systems}}
  \bibinfo{volume}{25}, \bibinfo{number}{10} (\bibinfo{year}{2017}),
  \bibinfo{pages}{2763--2775}.
\newblock


\bibitem[\protect\citeauthoryear{Wu, Yu, et~al\mbox{.}}{Wu
  et~al\mbox{.}}{2015}]%
        {wu2015deep}
\bibfield{author}{\bibinfo{person}{Jiajun Wu}, \bibinfo{person}{Yinan Yu},
  {et~al\mbox{.}}} \bibinfo{year}{2015}\natexlab{}.
\newblock \showarticletitle{Deep multiple instance learning for image
  classification and auto-annotation}. In \bibinfo{booktitle}{\emph{CVPR}}.
  \bibinfo{pages}{3460--3469}.
\newblock


\bibitem[\protect\citeauthoryear{Ye, Wang, et~al\mbox{.}}{Ye
  et~al\mbox{.}}{2018}]%
        {ye2018learning}
\bibfield{author}{\bibinfo{person}{Jinmian Ye}, \bibinfo{person}{Linnan Wang},
  {et~al\mbox{.}}} \bibinfo{year}{2018}\natexlab{}.
\newblock \showarticletitle{Learning compact recurrent neural networks with
  block-term tensor decomposition}. In \bibinfo{booktitle}{\emph{CVPR}}.
  \bibinfo{publisher}{IEEE}, \bibinfo{pages}{9378--9387}.
\newblock


\bibitem[\protect\citeauthoryear{Zhang, Lei, et~al\mbox{.}}{Zhang
  et~al\mbox{.}}{2018}]%
        {zhang2018stabilizing}
\bibfield{author}{\bibinfo{person}{Jiong Zhang}, \bibinfo{person}{Qi Lei},
  {et~al\mbox{.}}} \bibinfo{year}{2018}\natexlab{}.
\newblock \showarticletitle{Stabilizing gradients for deep neural networks via
  efficient SVD parameterization}.
\newblock \bibinfo{journal}{\emph{arXiv preprint arXiv:1803.09327}}
  (\bibinfo{year}{2018}).
\newblock


\end{thebibliography}
